\documentclass[epj]{svjour}
\usepackage{graphics}
\usepackage{placeins}
\usepackage{graphicx}
\usepackage[square,numbers]{natbib}

\newcommand{\e}{\mathrm{e}}
\DeclareMathAlphabet{\pazocal}{OMS}{zplm}{m}{n}
\begin{document}
\title{Trapping of null geodesics in slowly rotating extremely compact\\ Tolman VII spacetimes}
\author{Zden\v{e}k Stuchl\'{\i}k\inst{1}%
\thanks{\emph{Email address:} zdenek.stuchlik@physics.slu.cz}
\and Jaroslav Vrba\inst{1}
\thanks{\emph{Email address:} jaroslav.vrba@physics.slu.cz}%
}                     
%
%
\institute{$^\textrm{1}$Research Centre for Theoretical Physics and Astrophysics, Institute of Physics, Silesian University in Opava, Bezru\v{c}ovo n\'am.~13, 746\,01, Opava, CZ }
\date{Received: date / Revised version: date}
%
\abstract{
Region of trapped null geodesics hidden inside of extremely compact objects is of astrophysical importance because of trapping of gravitational waves, or neutrinos. The trapping effect of null geodesics was extensively studied for spherically symmetric extremely compact objects. Recently, influence of rotation of the extremely compact objects on the trapping of null geodesics was treated in the simplest possible model of internal linearised \mbox{Hartle--Thorne} spacetime with uniform energy density distribution and uniform emissivity distribution of null geodesics. Here we extend the study of the rotation influence on the trapping effect in the case of linearized \mbox{Hartle--Thorne} spacetimes based on the Tolman VII spherically symmetric solutions, where we assume the emissivity of the null geodesics proportional to the energy density of the Tolman VII object having quadratic radial profile. We demonstrate enhancement (suppression) of the trapping effect in the case of counter-rotating (co-rotating) null geodesics due to the behavior of the effective potentials and escape cones of the null geodesics in the linearised Hartle--Thorne--Tolman VII spacetimes. In dependence on the parameters of these spacetimes we determine the ``local'' and ``global'' coefficients of efficiency of the trapping and compare the results to those related to the rotating spacetimes based on the internal Schwarzschild spacetimes. We demonstrate that in the Tolman VII spacetimes the trapping is more efficient, being allowed in objects with radii larger than those of the trapping internal Schwarzschild spacetimes, occurring even for $R>3.3M$.
\PACS{
      {PACS-key}{discribing text of that key}   \and
      {PACS-key}{discribing text of that key}
     } 
} 
\maketitle
\section{Introduction}

The extremely compact objects allow, by definition, for existence of region of trapped null geodesics \cite{Fel:1968:NCBS,Abr-Mil-Stu:1993:PRD} that can be reflected due to trapping of gravitational waves \cite{Abr-And-Bru:1997:CQG} or neutrinos \cite{Stu-Tor-Hle:2009:CQG} -- we can thus denote such objects for short as trapping compact objects. The trapping region is centered around a stable circular null geodesic and limited by related unstable circular null geodesic \cite{Stu-Tor-Hle:2009:CQG}. The trapping compact objects can serve as mimickers of black holes, so their astrophysical relevance increases in connection with detection of gravitational waves due to merging black holes, their mimickers, or neutron stars \cite{Bar-etal:2019:CQG}. We expect connection of the quasinormal modes of gravitational waves and unstable circular null geodesics that was demonstrated for the objects described by the standard general relativity \cite{Car-Mir-Ber:2009:PRD}, but important exceptions of this expectation can exist \cite{Kon-Stu:2017:PLB,Tos-Stu-Sch:2018:PRD,Stu-Sch:2019:EPJC,Tos-Stu-Ahm:2019:GRQC}. 

The null geodesics inside the trapping compact objects representing neutron (quark) stars can govern motion of neutrinos, if the neutron starts are sufficiently cooled -- considering the neutrino scatter on electrons and neutrons, the limiting temperature is about $T_{geo} \sim 10^{9}K$ that is reached one day after the creation of the neutron star -- for details see \cite{Gle:2000:BOOK,Web:1999:BOOK,Stu-Tor-Hle:2009:CQG,Sha-Stu-Teu:1986:book}. The neutrino trapping can be in such cooled neutron stars important for several reasons. First, it decreases the neutrino flow observed at large distances during initial stages of a neutron star existence. Second, the trapped neutrinos significantly influence cooling of the neutron star -- its internal structure can be modified due to induced internal flows, causing self-organization of the neutron star matter in the trapping region and its surrounding, as mentioned in the study of the internal Schwarzschild spacetime \cite{Stu-Tor-Hle:2009:CQG}, or of its modification caused by the presence of the cosmological constant \cite{Stu-Hla-Urb:2011:GRG}; for relevance of the cosmological constant in astrophysical phenomena see \cite{Stu:2005:MPLA,Stu-Kol-Kov:2020:UNI}. 

The studies of the trapping compact objects were till now based on spherically symmetric spacetimes representing non-vacuum solutions of general relativity, or an alternative gravity. In the case of the internal Schwarzschild spacetimes with uniform distribution of energy density \cite{Stu-Tor-Hle:2009:CQG} it has been explicitly demonstrated that the trapping can occur only if the radius of the sphere $R<3GM/c^2=3r_\mathrm{g}/2$, where M is the mass and $r_\mathrm{g}$ is the related gravitational radius of the spacetime. Matter of such trapping compact objects must be thus concentrated in the region smaller than the radius of the unstable null circular geodesic (photosphere) of the external vacuum Schwarzschild spacetime. Efficiency of the neutrino trapping increases with decreasing radius of the uniform sphere, with $R\geq R_\mathrm{c}$ where $R_\mathrm{c}=9r_\mathrm{g}/8$ is the minimal value allowed for the radius of the internal Schwarzschild spacetimes \cite{Stu-Tor-Hle:2009:CQG}. \footnote{The special character of the internal Schwarzschild spacetimes in the limit of $R \to 2M$ related to gravastarts is discussed in \cite{Kon-Pos-Stu:2019:PRD,Pos-Chi:2019:CQG,Ova-Pos-Stu:2019:CQG}.} Trapping of neutrinos in the braneworld extremely compact objects, governed by an alternative gravity theory \cite{Ran-Sun:1999:PRL}, demonstrates similar consequences as shown in \cite{Stu-Hla-Urb:2011:GRG}.

An important question then arises, if the trapping compact objects can exist having their radius overcoming the radius of the photosphere, closer to the radii of real neutron stars that are observed larger than $R=3.2M$ \cite{Stu-Vrb-Hla:2021:sub1}. This requirement is satisfied by the Tolman VII solution of the Einstein gravitational equations that simply generalizes the internal Schwarzschild solution, assuming in the object a very simple, but non-uniform, energy density radial profile of quadratic character \cite{Tol:1939:PR}. The Tolman VII solution demonstrates trapping versions with radii reaching $R_\mathrm{tr}=3.202M$, significantly overcoming the radius of the photosphere \cite{Nea-Ish-Lak:2001:PRD}, being thus more plausible from the point of view of astrophysics in comparison with the trapping spacetimes limited by the photosphere radius. \footnote{The extremely compact polytropic spheres can also have extension overcoming the Schwarzschild photosphere radius $R=3M$ \cite{Stu-Hle-Nov:2016:PRD,Nov-Hla-Stu:2017:PRD,Stu-Sch-Tos:2017:JCAP,Hod:2018:PRD,Pen:2020:ARX}.} 

Properties of the Tolman VII solution \cite{Nea-Ish-Lak:2001:PRD,Jia-Yag:2019:PRD,Jia-Yag:2020:PRD} demonstrate that this exact solution of the Einstein equations exhibits surprisingly good approximation to properties of realistic neutron stars. The modified Tolman VII solution introduced in \cite{Jia-Yag:2019:PRD} includes an additional quartic term in the energy density radial profile being an approximate solution of the Einstein equations. Its concordance with realistic models of neutron stars was confirmed due to the I-Love-C relation \cite{Jia-Yag:2020:PRD} and the dynamic stability \cite{Pos-Hla-Stu:2021:sub}. Furthermore, anisotropic version of the Tolman VII solution \cite{Hen-Stu:2019:GRQC} enables inclusion of the influence of additional matter sources on the properties of neutron stars. 

The neutrino trapping effect in the exact Tolman VII solution of the Einstein equations, giving good approximation of the realistic neutron stars, has been recently treated in \cite{Stu-Vrb-Hla:2021:sub1}. Here we extend study of the trapping Tolman VII spacetimes for the influence of slow rotation of such spheres. 

The influence of the spacetime rotation on the trapping of null geodesics was treated for the first time in the simplest approximation of rotating compact objects by the linearised Hartle--Thorne model having uniform distribution of energy density of the matter content. The standard internal Schwarzschild spacetime with the uniform energy density distribution, enriched by the Lense -- Thirring off-diagonal term governed by the Hartle--Thorne equation for the dragging of inertial frames \cite{Har-Tho:1968:ApJ}, was used -- uniformly distributed and isotropically and uniformly radiating sources of null geodesics (neutrinos) were assumed \cite{Vrb-Urb-Stu:2020:EPJC}. Surprisingly, even on the level of the linearized model, a strong asymmetry of the trapping of the counter-rotating (enhancement) and co-rotating (suppression) null geodesics has been demonstrated, along with strong dependence on the compactness of the sphere, and the angular velocity of its rotation \cite{Vrb-Urb-Stu:2020:EPJC}. Here we generalize the study of the role of the rotation on the trapping effect for the case of linearized Hartle--Thorne--Tolman VI model, assuming radius dependent distribution of the sources of the null geodesics, namely given by the distribution of the energy density in the Tolman VII solution. We first introduce the spherically symmetric trapping Tolman VII spacetimes and shortly review the null geodesic trapping in these spacetimes. Then we introduce the slowly rotating Tolman VII spacetime in the linear approximation of the Hartle--Thorne theory, and define the effective potential and escape (trapping) cones of their null geodesics. We define and calculate the \lq local\rq \ and \lq global\rq \ efficiency of trapping of null geodesics in rotating Tolman VII spacetime for different values of compactness, rotation, and position in the object and compare the trapping efficiencies of rotating and non-rotating configurations. Finally, we determine the "local" and "global" efficiency coefficients of the trapping effect, and compare the results to those related to the slowly rotating internal Schwarzschild spacetimes.

\section{Trapping Tolman VII spacetime and its null geodesics}

As the starting point of our study we consider the Tolman VII spacetime \cite{Tol:1939:PR} representing direct exact static and spherically symmetric generalization of the simplest internal solution of the Einstein gravitational equations for uniform distribution of the energy density of compact objects given by the internal Schwarzschild spacetime \cite{Sch:1916:AkadW,Stu:2000:APS,Boh:2004:GRG}. In the  Tolman VII solution, governing the object composed from perfect fluid with energy density $\rho$ and pressure $p$, the energy density distribution is assumed to be quadratic function of the radius what enables determination of the metric coefficients and pressure radial profiles in terms of elementary functions\cite{Jia-Yag:2019:PRD}. 

We first present the Tolman VII solution \cite{Tol:1939:PR} in an elegant and compact form introduced in \cite{Jia-Yag:2019:PRD}, using as free parameter the compactness of the object $\pazocal{C}=M/R$, or its inverse, $R/M$, where $M$ is the mass of the object and $R$ is its radius. Then we determine the values of the parameter $R/M$ corresponding to the trapping (extremely compact) Tolman VII solutions due to the behavior of the null geodesics of the spacetime. 

\subsection{Tolman VII spacetimes}

The static and spherically symmetric spacetimes are in the standard Schwarzschild coordinates ($t,r,\theta,\varphi$) given by the line element taking the form 
\begin{eqnarray}
&&\mathrm{d}s^2=-e^{\Phi(r)}\, \mathrm{d}t^2+e^{\Psi(r)}\, \mathrm{d}r^2 +r^2\big(\mathrm{d}\theta^2 +
\sin^2\theta\, \mathrm{d}\varphi^2\big).\quad
\label{met} 
\end{eqnarray}
The exterior of the Tolman VII object, $r>R$, is governed by the external vacuum Schwarzschild geometry with coefficients
\begin{equation}
g_{tt}=-e^{\Phi}=-\Bigg(1-\frac{2M}{r} \Bigg),\, g_{rr}=e^{\Psi}=\Bigg(1-\frac{2M}{r} \Bigg)^{-1}.
\end{equation}

The interior of the Tolman VII solution is determined by assumption on the behavior of the energy density radial profile \cite{Tol:1939:PR} that is given in terms of the dimensionless radial coordinate $\xi = \frac{r}{R}$ by the relation 
\begin{equation}
\rho(\xi) = \rho_{c}(1 - \xi^2) , 
\end{equation} 
$\rho_c$ denotes central energy density. The Einstein gravitational equations immediately imply \cite{Jia-Yag:2019:PRD} the mass radial profile in the form 
\begin{equation}
m(r) = 4\pi\rho_c\left(\frac{r^3}{3} - \frac{r^5}{5R^2}\right)  
\end{equation} 
that enables to give the central density in terms of the compactness parameter $\pazocal{C}$ as    
\begin{equation}
\rho_c = \frac{15\pazocal{C}}{8\pi R^2} . 
\end{equation} 
The Tolman VII solution can be then expressed in the simple form introduced by \cite{Jia-Yag:2019:PRD,Jia-Yag:2020:PRD}. The energy density and mass profiles read
\begin{equation}
\rho(\xi) = \frac{15\pazocal{C}}{8\pi R^2}(1 - \xi^2) , 
\end{equation} 
\begin{equation}
m(\xi) = \pazocal{C}R\left(\frac{5}{2}\xi^3 - \frac{3}{2}\xi^5\right) ,  
\end{equation} 
while the metric coefficients can be determined in the form  
\begin{equation}
e^{-\Psi(\xi)} = 1 - \pazocal{C}\xi^2(5 - 3\xi^2) ,  
\end{equation} 
\begin{equation}
e^{\Phi(\xi)} = C_1 \cos^2\phi_T , 
\end{equation} 
and the pressure radial profile is given by the relation  
\begin{equation}
p(\xi) = \frac{1}{4\pi R^2}\left[\sqrt{3\pazocal{C}e^{-\Psi}}\tan\phi_T - \frac{\pazocal{C}}{2}(5 - 3\xi^2)\right]
\end{equation} 
where
\begin{equation}
\phi_{T}(\xi) = C_2 - \frac{1}{2}\log\left(\xi^2 - \frac{5}{6} + \sqrt\frac{e^{-\Psi}}{3\pazocal{C}}\right)
\end{equation} 
and the integration constants are given as  
\begin{equation}
C_1 = 1 - \frac{5\pazocal{C}}{3} , 
\end{equation} 
\begin{equation}
C_2 = \arctan\sqrt\frac{\pazocal{C}}{3(1-2\pazocal{C})} + \frac{1}{2}\log\left(\frac{1}{6} + \sqrt\frac{1 - 2\pazocal{C}}{3\pazocal{C}}\right) . 
\end{equation} 

Existence of the Tolman VII spacetimes is limited by the requirement of finite central pressure. The central pressure $p(\xi=0)$ diverges, if $\tan\phi_{T}(\xi=0)$ diverges, i.e., if 
\begin{equation}
	\phi_{T}(\xi=0)=\frac{\pi}{2} . 
\end{equation}
We thus arrive to the relation 
\begin{equation}
  \frac{\pi}{2} = C_2 - \frac{1}{2} \log(\sqrt{\frac{1}{3\pazocal{C}}} - \frac{5}{6})  
\end{equation}
implying the lower limiting value of the Tolman VII sphere 
\begin{equation} 
                  R_\mathrm{Tmin}\doteq 2.589M .
\end{equation}
The minimal radius of the Tolman VII solutions exceeds significantly the lowest radius of the internal Schwarzschild solutions that reads $R_\mathrm{Smin} =  2.25M$ \cite{Abr-Mil-Stu:1993:PRD,Stu-Tor-Hle:2009:CQG}. Clearly, $R_\mathrm{min}$ is also the lowest radius of the extremely compact, trapping Tolman VII solutions. The upper limit on the trapping Tolman VII spacetimes is determined by properties of the null geodesics of these spacetimes. 

\subsection{Effective potential of the null geodesic and trapping Tolman VII spacetimes}

The null geodesics are governed by the geodesics equation and related norm condition for the 4-momentum $p^{\mu}$
\begin{equation}
\frac{\mathrm{D}p^{\mu}}{\mathrm{d}\tau}=0,\qquad p_{\mu}p^{\mu}=0,
\label{geod}
\end{equation}
where $\tau$ is the affine parameter. Projections of the 4-momentum $p^{\mu}$ onto two Killing vector fields $\partial/\partial t\ \mathrm{and}\  \partial/\partial \varphi$  imply two conserved components of the 4-momentum:
\begin{equation}
p_t=-E\quad \mathrm{(energy)},\quad p_{\varphi}=\phi\quad \textrm{(axial angular momentum)}.
\label{const}
\end{equation}
In the spherically symmetric spacetimes the geodesic motion is always restricted to the central planes, and it is convenient to choose the equatorial plane with $\theta=\pi/2=$ constant, if one geodesic is considered. The equation of the radial motion can be then obtained from the normalization condition that takes, after introducing the impact parameter $\lambda=\phi/E$, the form \cite{Mis-Tho-Whe:1973:BOOK}
\begin{equation} 
(p^r)^2=\frac{-1}{g_{tt}g_{rr}}E^2\Bigg(1+g_{tt}\frac{\lambda^2}{r^2}\Bigg).
\end{equation}
Clearly, the energy $E$ is not substantial and can be used to scale the affine parameter $\tau$. The turning points of the radial motion along a null geodesic with a given impact parameter $\lambda$ can be thus governed by the effective potential introduced in the following way: 
\begin{equation}
\lambda^2\leq \mathrm{V}_\mathrm{eff}=\cases{ \mathrm{V}_\mathrm{eff}^\mathrm{int}=\frac{3r^2}{3-5\pazocal{C}}\cos ^{-2}\left(C_a+Y(r)\right) \quad \textrm{for } r \le R\\
	\mathrm{V}_\mathrm{eff}^\mathrm{ext}=\frac{r^3}{r-2M}\quad \textrm{for } r > R,}
\label{e:sveff}
\end{equation}

where 
\begin{eqnarray}
&&C_a=\arctan\sqrt{\frac{\pazocal{C}}{3-6 \pazocal{C}}}, \nonumber \\
&&Y(r)=\frac{1}{2}\log\left[\frac{\left(1 + 2 \sqrt{\frac{3}{\pazocal{C}} - 6}\right) R^2}{2 \sqrt{\frac{3R^4}{\pazocal{C}} + 9 r^4 - 15 r^2R^2}+6 r^2 - 5R^2}\right].
\label{e:helpc}
\end{eqnarray}
\begin{figure}[ht]
	\centering
	\includegraphics[width=0.7\hsize]{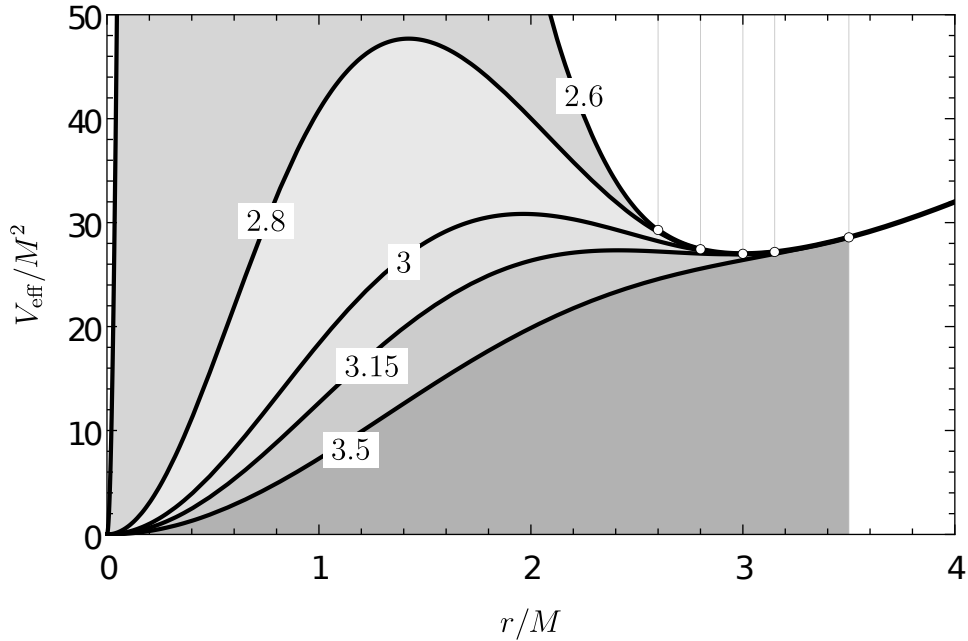}
	\caption{The effective potential of null geodesics given for typical Tolman VII objects with $R/M=\{2.6,\ 2.8,\ 3,\ 3.15,\ 3.5\}$. With exception of the case $R/M=3.5$, where the effective potential has no extrema, all the other Tolman VII objects are trapping, with two local extrema of the effective potential.}
	\label{f:efpot}
\end{figure}

The resulting effective potentials are illustrated in Fig.~\ref{f:efpot}; notice that contrary to the case of the internal Schwarzschild spacetimes, the effective potential of the Tolman VII spacetimes can be non-monotonic for $R>3M$. 

The circular null geodesics are determined by the condition $\mathrm{d}V_\mathrm{eff}^\mathrm{int}/\mathrm{d}r=0$ implying the relation 
\begin{eqnarray}
\frac{6 r \left[r Y'(r) \tan \chi+1\right]}{(5 \pazocal{C}-3)\cos ^{2}\chi}=0, 
\label{e:dVef}
\end{eqnarray}
where
\begin{equation}
   \chi = C_a + Y(r).
\end{equation}
The limiting case corresponds to the inflexion point of the effective potential, when the additional condition $\mathrm{d}^2V_\mathrm{eff}^\mathrm{int}/\mathrm{d}r^2=0$ implying the relation 
\begin{eqnarray}
\frac{r\left[ 2 r Y'(r)^2 \left(\cos (2 \chi)-2\right)-r Y''(r) \sin (2 \chi)-4 Y'(r) \sin (2 \chi)\right]}{\cos^{2}{\chi}}=2  
\end{eqnarray}
has to be added. These conditions give the critical maximal value of $R$ allowing for existence of the trapping Tolman VII solutions 
\begin{equation}  
                 R_t \doteq 3.202 M.  
\end{equation}

\begin{figure}[ht]
	\centering
	\includegraphics[width=1\hsize]{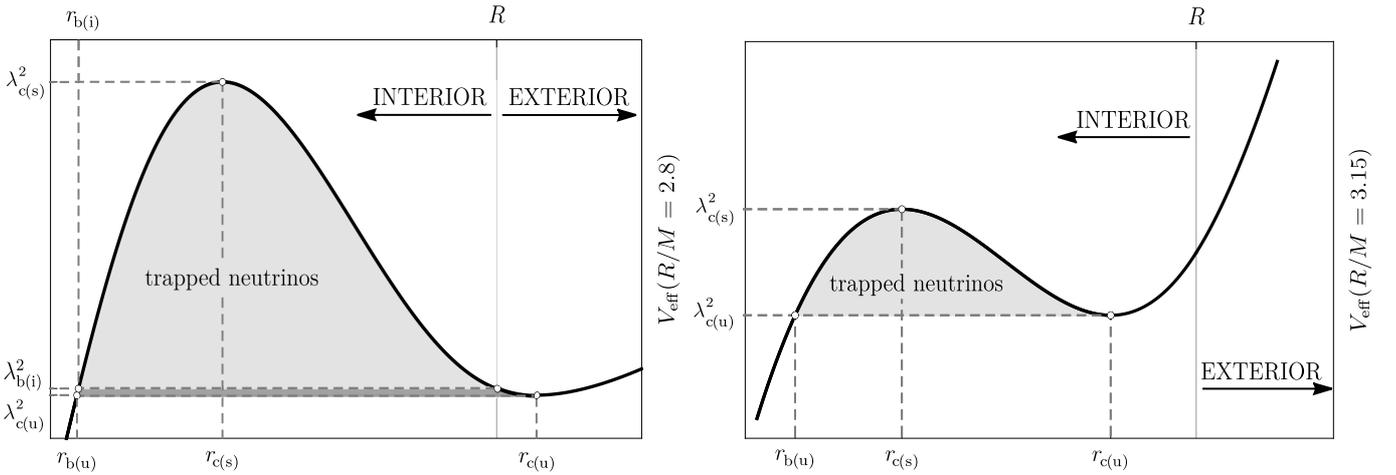}
	\caption{The effective potentials for an object
		with $R=2.8M$ (left panel) and $R=3.15M$ (right panel); the trapped areas are shaded. The lightly-shaded 
		area, bounded by the radii $r_\mathrm{b(u)}$ and $R$, contains null 
		geodesics trapped completely inside the object, while the 
		darkly-shaded area, bounded by the radii $r_\mathrm{b(u)}$ and 
		$r_\mathrm{c(u)}$, contains null geodesics, which can temporarily pass 
		through the surface into the vacuum before returning.}
	\label{f:vef}
\end{figure}

The trapping Tolman VII spacetimes thus exist for $R_\mathrm{Tmin}/M \leq R/M \leq R_\mathrm{t}/M$. They can be separated into two classes. For the first class with $R/M \leq 3$, the unstable circular null geodesic exists in the vacuum Schwarzschild spacetime, beeing related to the local minimum of the external effective potential, $V_\mathrm{eff}^\mathrm{ext}$.
The effective potential of the Tolman VII spacetime has then similar character to those of the internal Schwarzschild spacetimes, as demonstrated in Fig.~\ref{f:vef} (left). The (shaded) region of trapped null geodesics consists of two parts -- the internal trapped null geodesics (dark shade) are fully restricted to the interior of the object,  while the external trapped null geodesics (light shade) leave and re-enter the object, see \cite{Stu-Tor-Hle:2009:CQG}. 

In the second class with $3M<R/M<R_\mathrm{t}/M$, the internal effective potential $V_\mathrm{eff}^\mathrm{int}(r)$ of the Tolman VII spacetime demonstrates along with presence of the local maximum giving the stable circular null geodesics also a local minimum giving an unstable circular null geodesic and the motion along all of the trapped null geodesics is fully restricted to the interior as shown in Fig.~\ref{f:vef} (right). 

Detailed discussion of the properties of these trapping spacetimes can be found in \cite{Stu-Vrb-Hla:2021:sub1}. 

\FloatBarrier

\subsection{Escape cones and trapped null geodesics}

Trapping of null geodesics in the context of neutrinos radiated by matter of the trapping compact objects is determined by the trapping cones of null geodesics (complementary to the escape cones) related to the radiating matter of the object - the trapped neutrinos are radiated in angles lying outside of the escape cone. We first summarize establishing of the trapping cones in the spherically symmetric Tolman VII spacetimes as presented in \cite{Stu-Vrb-Hla:2021:sub1}. 

The escape (trapping) cones have to be related to the static sources in these static spacetimes and have to be connected to the corresponding local frames of the static observers. The tetrad of differential forms in the static and spherically symmetric spacetimes is defined as
\begin{eqnarray}
e^{(t)}=\mathrm \e^{\Phi/2}\, \mathrm d t,~~  e^{(r)}= \e^{\Psi/2}\, \mathrm d r,~~ e^{(\theta)}=r\, \mathrm d \theta,~~ e^{(\varphi)}=r\, \mathrm{sin}\theta\, \mathrm d \varphi.
\end{eqnarray}
The complementary tetrad of 4-vectors $e_{(\alpha)}$, is given by
\begin{eqnarray}
e_{(\alpha)}^{\mu} e_\nu^{(\alpha)}=\delta^\mu_\nu,~~ e_\mu^{(\alpha)} e^\mu_{(\beta)}=\delta^\alpha_\beta.
\end{eqnarray}
Projections of a 4-momentum $p^{\mu}$ take the form
\begin{eqnarray}
p^{(\alpha)}=p^{\mu} e_\mu^{(\alpha)},~~p_{(\alpha)}=p_\mu e^\mu_{(\alpha)}.
\end{eqnarray}

\begin{figure}[ht]
\centering
\includegraphics[width=0.5\hsize]{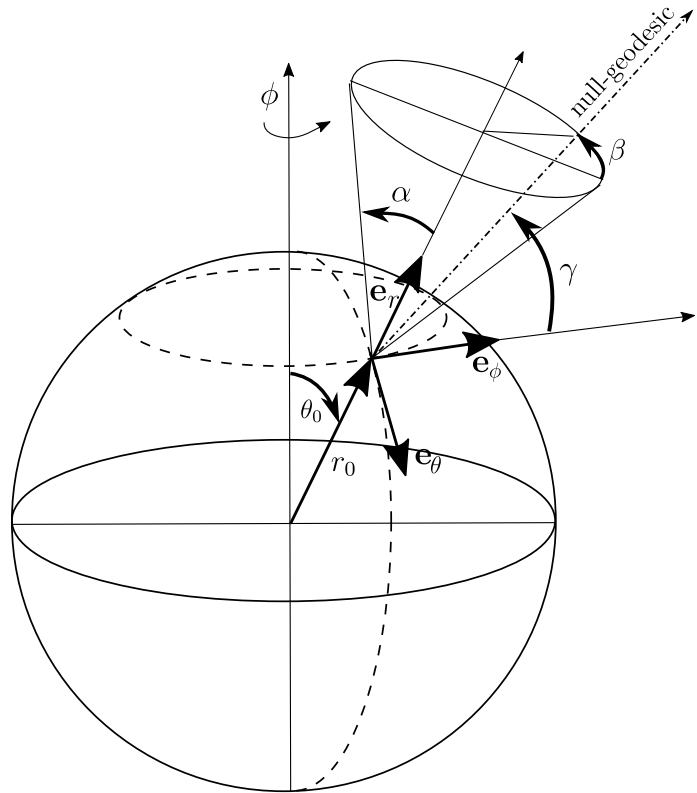}
\caption{The definition of directional angles governing direction of a null geodesic at a given point of the spacetime. The angles are related to the outward radial direction.}
\label{f:defangles}
\end{figure}

The neutrinos radiated by a static source can be described by the directional angles defined in Fig.~\ref{f:defangles}, being related to the outgoing radial direction in the spacetime. The angles {$\alpha,~\beta,~\gamma$} are connected by the relation 
\begin{equation}
\cos\gamma=\sin\beta\, \sin\alpha.
\end{equation}
The angle $\alpha$ is taken between the null geodesic direction and the outgoing radial unit vector, the angle $\beta$ determines position of the cone. The angle $\gamma$ is taken betwen the null geodesic direction and the unit vector in the azimuthal direction. Because of the spherical symmetry of the spacetime, the ecape (trapping) cone is fully determined by the angle $\alpha$. 

The trapping (escape) cone in the observer (source) sky is determined by the angles corresponding to null geodesic with parameter defining the unstable null circular geodesics and the one related to the radius where the observer is located. The maximal extension of the trapping cone occurs at the radius of stable circular null geodesic and corresponds to its impact parameter. At given radius the escape (trapping) cone is defined by the extension of the effective potential barrier where the trapping occurs. In order to determine the maximal extension of the trapping cone we have to find the angles $\alpha_\mathrm{c(s)}$, corresponding to $\lambda_\mathrm{c(s)}$ of the stable null  circular geodesic, and $\alpha_\mathrm{c(u)}$, corresponding to the impact parameter $\lambda_\mathrm{c(u)}$ of the unstable null circular geodesic.

Therefore, we relate the directional angles to the impact parameters of the null geodesics. The spherical symmetry enables to restrict attention to the equatorial null geodesics (when $\beta=0$, or $\beta=\pi$). The directional angle $\alpha$ is then governed by the relations ($p^{(\theta)}=0$)
\begin{equation}
\sin\alpha=\frac{p^{(\varphi)}}{p^{(t)}},~~\cos\alpha=\frac{p^{(r)}}{p^{(t)}}.
\end{equation}

Because the radial component of the 4-momentum can be expressed as 
\begin{equation}
p_r=\pm E \mathrm e^{(\Psi-\Phi)/2}\left(1-\mathrm e^{\Phi}\frac{\lambda^2}{r^2}\right)^{1/2},
\end{equation}
we give the directional angle corresponding to a given impact parameter $\lambda$ in the form 
\begin{eqnarray}
\sin\alpha&=&\e^{\Phi/2}\,\frac{\lambda}{r} = \sqrt{1-5\pazocal{C}/3}\, \cos\left[C_\mathrm{a} + Y(r)\right] \,\frac{\lambda}{r},\\
\cos\alpha&=&\pm\sqrt{1-\sin^2\alpha} . 
\end{eqnarray}

Using the relations for directional angle with $\lambda=\lambda_\mathrm{c(s)}$ and $\lambda=\lambda_\mathrm{c(u)}$, we arrive to searched critical angles determining the maximal trapping in the form  
\begin{eqnarray}
\cos\alpha_\mathrm{c(s)}(r,\pazocal{C})&=&\pm \sqrt{1 - \left( \e^{\Phi(r)/2}\, \frac{\lambda_\mathrm{c(s)}}{r}\right)^2},\\
\cos\alpha_\mathrm{c(u)}(r,\pazocal{C})&=&\pm \sqrt{1 - \left( \e^{\Phi(r)/2}\frac{\lambda_\mathrm{c(u)}}{r}\right)^2}.
\end{eqnarray}
For trapping given by the external Schwarzschild geometry there is 
\begin{equation}
\sin\alpha_\mathrm{c(u)}(r,R)=\e^{\Phi(r)/2}\, \frac{3\sqrt{3}}{r}.
\end{equation}

\begin{figure}[ht]
\centering
\includegraphics[width=0.40\hsize]{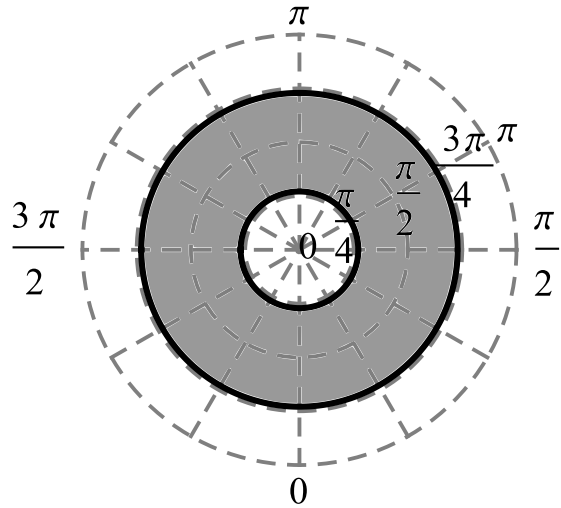}
\caption{An example of a maximal trapping cone for non-rotating configuration with $R/M=2.8$ at the position of stable circular null geodesic.}
\label{scone}
\end{figure}

There is always $\alpha_\mathrm{c(s)}>\alpha_\mathrm{c(u)}$, and the trapping zone lies between these angles, as shown in Fig.~\ref{scone} where the maximal trapping zone is given. The escape cone (zone) is light and null geodesics of this kind can escape to infinity even if originally radiated inwards \footnote{Note that in the rotating spacetimes the symmetry of the trapping zone (cone) is lost as the motion depends on the sign of the impact parameter, as demonstrated for the case of trapped null geodesics in Kerr spacetimes \cite{Stu-Sch:2010:CQG,Stu-Cha-Sch:2018:EPJC}}. 
Recall that the trapping effects are relevant only in the extremely compact Tolman VII spacetimes existing for the inverse compactness in the range $2.588 < R/M < 3.202$, and can occur only in the range of radii of these objects limited by $ r_\mathrm{b(u)} < r < r_\mathrm{c(u)}$ for $R/M>3$, and $ r_\mathrm{b(u)} < r < R$ for $R/M<3$. 

The trapping cones were in the case of the spherically symmetric Tolman VII spacetime used to determine efficiency of the trapping of null geodesics (and estimate the neutrino trapping) in these spacetimes \cite{Stu-Vrb-Hla:2021:sub1}. Here we study the trapping phenomena for linearised Hartle--Thorne models based on the Tolman VII spacetimes as the background solution, endowed by the linear rotational off-diagonal metric coefficient. The linearised Hartle--Thorne model corresponds to the Lense-Thirring metric at the exterior of the compact object \cite{Vrb-Urb-Stu:2020:EPJC}. 

\section{First-order Hartle--Thorne spacetime based on the Tolman VII solution}

The effect of rotation on the Tolman VII spacetime with quadratically distributed energy density will be treated in the framework of  Hartle--Thorne slow-rotation model considered to first order in the angular velocity of the object. In the framework of the Hartle--Thorne theory, both interior and exterior of relativistic perfect fluid rotating objects are considered up to the second order of the assumed uniform (rigid) rotation with angular velocity $\Omega$, as measured by distant static observers \cite{Har-Tho:1968:ApJ}. Here we restrict attention to the linearized form of the Hartle--Thorne model, neglecting in the metric all terms related to $\Omega^2$; the starting point is the Tolman VII solution, where all relevant quantities can be represented by elementary functions, similarly to the case of the internal Schwarzschild spacetimes discussed in \cite{Vrb-Urb-Stu:2020:EPJC}.

\subsection{Hartle--Thorne spacetimes and their locally non-rotating frames}

The internal Hartle--Thorne metric can be in the geometric units (c=G=1) written in the form using the 1-forms as 
\begin{equation}
\mathrm d s^2 = -\left[\mathrm e^{(t)} \right]^2 + \left[\mathrm e^{(r)} \right]^2 + \left[\mathrm e^{(\theta)} \right]^2 +\left[\mathrm e^{(\varphi)} \right]^2
\end{equation}
where the 1-forms are defined to reflect the locally non-rotating frames (LNRFs) \cite{Bar-Pre-Teu:1972:ApJ}, corresponding to observers orbiting at given $r, \theta$ with zero angular momentum (the term ZAMO frames is thus also used). The ZAMOs are locally stationary with respect to the internal Hartle--Thorne spacetime, being dragged to corotation with the angular velocity $\omega(r)$ characterizing the spacetime rotation relative to static distant observers: 
\begin{eqnarray}
&&\mathrm e^{(t)}=\sqrt{-g_{tt(\mathrm i)}} \mathrm d t, ~~\mathrm e^{(r)}=\sqrt{g_{rr(\mathrm i)}} \mathrm d r, \\ \nonumber
&&\mathrm e^{(\theta)}=\sqrt{g_{\theta\theta(\mathrm i)}} \mathrm d \theta, ~~
\mathrm e^{(\varphi)}=\sqrt{g_{\theta\theta(\mathrm i)}}\sin\theta(\mathrm d \varphi - \omega \mathrm d t).  
\end{eqnarray}
The metric coefficients are presented in detail in \cite{Har-Tho:1968:ApJ}, $\omega(r)$ is the angular velocity of the LNRFs reflecting the rotational dragging of the internal spacetime by the gravitation of the object rigidly rotating with the angular velocity $\Omega=$ const. The motion of the particles belonging to the for the first-order Hartle--Thorne--Tolman VII object corresponds to rigid and uniform rotation with angular velocity $\Omega$, as considered by distant static observers. Their \mbox{4-velocity} $u^\mu$ thus has components
\begin{eqnarray}
u^t=\left(-g_{tt} - 2\Omega g_{t\varphi} -g_{\varphi\varphi}\Omega^2 \right)^{-1/2}, ~~~
u^\varphi=\Omega u^t, u^r=u^\theta=0 . 
\end{eqnarray}

The relative angular velocity of the rotating matter to the rotating Hartle--Thorne spacetime 
\begin{equation}
\bar\omega(r) \equiv \Omega-\omega(r)
\end{equation}
enters the Einstein gravitational equations being of the first order in $\Omega$. It fulfills the differential equation 
\begin{equation}
\frac{1}{4}\frac{\mathrm{d}}{\mathrm{d}r}\Bigg(r^4f(r)\frac{\mathrm{d}\bar{\omega}}{\mathrm{d}r}
\Bigg)+\frac{4}{r}\frac{\mathrm{d}f(r)}{\mathrm{d}r}\bar{\omega}=0,
\label{difo}
\end{equation}
where 
\begin{equation}
f(r)=\Bigg(\frac{-1}{g_{tt(i)}g_{rr(i)}}\Bigg)^{1/2}.
\label{e:f}
\end{equation}
The metric coefficients $g_{tt(i)}$ and $g_{rr(i)}$ are given by the internal static spherically symmetric spacetime considered as basic for the calculations. The solution of this equation is crucial in the framework of the Hartle--Thorne theory and serves equally for both the first-order and second-order models. It has to be regular at the origin $r=0$ -- namely d$\bar\omega/$d$r=0$; the angular velocity $\bar\omega$ reaches at the origin a finite value $\bar\omega_c$, governed by the matching with the exterior vacuum solution at $r=R$. The matching procedure implies expressions for the angular momentum $J$ of the object and the angular velocity $\Omega$ of the object \cite{Har-Tho:1968:ApJ} taking the form 
\begin{equation}
J=\frac{1}{6}R^4\Bigg(\frac{\mathrm{d}\bar{\omega}}{\mathrm{d}r} \Bigg)_{r=R},
\end{equation}
and 
\begin{equation}
\Omega=\bar\omega(R) + \frac{2J}{R^3}.
\end{equation}
It is convenient to solve the rotation equation for $\bar\omega(r)$ related to $\bar\omega_c$, then find $J/\bar\omega_c$ and $\Omega/\bar\omega_c$, and finally to express $J$ and $\bar\omega(r)$ in units of $\Omega$ \cite{Mil:1977:MNRAS}. 
Then we can find the moment of inertia $I=J/\Omega$ and the radius of gyration (see \cite{Abr-Mil-Stu:1993:PRD})
\begin{equation}
\mathcal R_\mathrm{gyr}=\left( \frac{J}{\Omega M} \right)^{1/2}.
\end{equation}

The exterior of the object can be described by the external Hartle--Thorne geometry that can be expressed in the tetrad frame taking the form \cite{Har-Tho:1968:ApJ}
\begin{eqnarray}
&&\mathrm e^{(t)}=\sqrt{-g_{tt(\mathrm e)}} \mathrm d t, ~~\mathrm e^{(r)}=\sqrt{g_{rr(\mathrm e)}} \mathrm d r, \nonumber\\ 
&&\mathrm e^{(\theta)}=\sqrt{g_{\theta\theta(\mathrm e)}} \mathrm d \theta, ~~
\mathrm e^{(\varphi)}=\sqrt{g_{\theta\theta(\mathrm e)}}\sin\theta(\mathrm d \varphi - \frac{2J}{r^3} \mathrm d t).
\end{eqnarray}
Exact form of the metric coefficients of the external spacetime can be found in \cite{Har-Tho:1968:ApJ}, and in an alternative form in \cite{Abr-Alm-Klu:2003:arx}.  

\subsection{First-order Hartle--Thorne--Tolman VII spacetime}

\begin{figure}[ht]
	\centering
	\includegraphics[width=0.55\hsize]{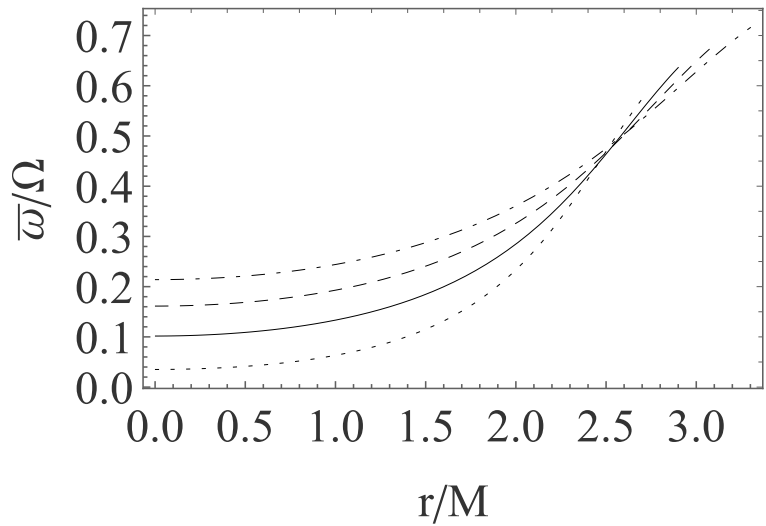}
	\caption{Numerical solution for $\bar{\omega}$
		scaled by $\Omega$ for $R/M=2.7$ (dotted line), 
		$R/M=2.9$ (solid line), $R/M=3.1$ (dashed line) and 
		$R/M=3.3$ (dash-dotted line).}
	\label{f:omega}
\end{figure}

The line element of the first-order (linearized) Hartle--Thorne internal spacetime related to the Tolman VII spherically symmetric solution takes in the geometric units ($c=G=1$) the form 
\begin{equation}
\mathrm{d}s^2=-e^{\Phi(r)}\,\mathrm{d}t^2+e^{\Psi(r)}\,\mathrm{d}r^2 +r^2\big(\mathrm{d}\theta^2 + \sin^2\theta\, \mathrm{d}\varphi^2\big) - 2 \omega(r)\, r^2 \sin^2\theta\, \mathrm{d}t\,\mathrm{d}\varphi,
\label{metLTi}
\end{equation} 
where $\Phi(r)$ and $\Psi(r)$ are the functions governing the static spherically symmetric Tolman VII configuration, and $\omega(r)$ is the angular velocity of the LNRF related to the first-order Hartle--Thorne--Tolman VII spacetimes. The first-order rotating spacetime geometry is therefore fully determined by finding the solution of Eq.~(\ref{difo}) for the angular velocity of the rotating matter relative to the spacetime geometry, $\bar\omega$, where the characteristic function of the solution related to the Tolman VII spacetimes reads
\begin{equation}
f(r)= \sqrt{\frac{1 - \pazocal{C}\left(r/R\right)^2\left[5 - 3\left(r/R\right)^2\right]}{C_1 \cos^2\left(\phi_T(r)\right)}} .
\end{equation}
The results of the integration are presented in Fig.~\ref{f:omega} -- we can see that the role of the rotational effects increases with decreasing of the inverse compactness of the object $R/M$, if the angular velocity $\Omega$ (or total angular momentum of the object $J$) are fixed. As in the case of the internal Schwarzschild spacetimes \cite{Vrb-Urb-Stu:2020:EPJC}, the strongest dragging effect ($\omega(r=0)=\Omega$) is obtained for the maximally compact spacetime with $R_\mathrm{tmin}=2.589M$. Of course, this is physically unattainable case, as it correspond to pressure divergent in the object center. The representative could be considered the four cases demonstrated in Fig.~\ref{f:omega}.

The metric of the external first-order Hartle--Thorne spacetime takes the standard Lense-Thirring form 

\begin{eqnarray}
\mathrm{d} s^2 = &-&\left(  1-\frac{2M}{r}   \right)  \mathrm{d}t^2+\left(  1-\frac{2M}{r}   \right)^{-1}\mathrm{d}r^2 \nonumber \\
 &+& r^2\left(\mathrm{d}\theta^2 + \sin^2\theta\, \mathrm{d}\varphi^2\right) - 4 \frac{J}{r} \sin^2\theta\, \mathrm{d}t\,\mathrm{d}\varphi . 
\label{metLTe}
\end{eqnarray} 

For our purposes, we need also the inverse form of the metric representing the first-order spacetimes. The contravariant form of the internal first-order Hartle--Thorne internal and external Tolman VII metric reads 
\begin{eqnarray}
          \left(\frac{\partial}{\partial s}\right)^{2}_{\mathrm{int}} = &&-e^{-\Phi(r)}_\mathrm{int} \frac{\partial^2}{\mathrm{\partial}t^2}+e^{-\Psi(r)}_\mathrm{int} \frac{\partial^2}{\mathrm{\partial}r^2} + \frac{1}{r^2}\left(\frac{\partial^2}{\mathrm{\partial}\theta^2} + \frac{1}{\sin^2\theta} \frac{\partial^2}{\mathrm{\partial}\varphi^2}\right)\nonumber \\
            &&-\omega(r)e^{-\Phi(r)}_\mathrm{int} \frac{\partial^2}{\mathrm{\partial}t\mathrm{\partial}\varphi},
\end{eqnarray}
where
\begin{eqnarray}
e^{-\Phi(r)}_\mathrm{int}&=& \frac{3}{3-5\pazocal{C}}\cos ^{-2}\left(C_a+Y(r)\right) , \nonumber \\
e^{-\Psi(r)}_\mathrm{int}&=& 1 - \pazocal{C}\xi^2(5 - 3\xi^2)
\end{eqnarray}
and $C_a$, $Y(r)$ are defined in (\ref{e:helpc}). The external first-order metric takes the form 
\begin{eqnarray}
          \left(\frac{\partial}{\partial s}\right)^{2}_{\mathrm{ext}} =&& -\left(  1-\frac{2M}{r}   \right)^{-1}  \frac{\partial^2}{\mathrm{\partial}t^2}+\left(  1-\frac{2M}{r}   \right) \frac{\partial^2}{\mathrm{\partial}r^2} +\frac{1}{r^2}\left(\frac{\partial^2}{\mathrm{\partial}\theta^2} + \frac{1}{\sin^2\theta} \frac{\partial^2}{\mathrm{\partial}\varphi^2}\right)\nonumber \\
          && -\frac{4 J}{(r-2M) r^2} \mathrm{d}t\mathrm{d}\varphi.
\end{eqnarray}
The metric coefficient $g^{t\phi}$ is independent of the latitudinal coordinate $\theta$ due to the applied linear approximation, enabling thus separability of the equations of the geodesic motion in the first-order Hartle--Thorne spacetimes. This property simplifies substantially calculation of the trapping effect. 

In the first-order Hartle--Thorne internal spacetime the ZAMO observers have the four velocity 
\begin{equation}
               ( U^{\mu})_{\mathrm{ZAMO}} = e^{-\Phi(r)/2}(1,0,0,\omega),\quad (U_{\mu})_{\mathrm{ZAMO}} = e^{\Phi(r)/2}(-1,0,0,0)  
\end{equation}
and the LNRF tetrad of 1-forms now takes the form 
\begin{eqnarray}
    && e^{(t)}_{\mu} = e^{\Phi(r)/2}(1,0,0,0),\nonumber\\  
    && e^{(r)}_{\mu} = e^{\Psi(r)/2}(0,1,0,0),  \nonumber\\
    && e^{(\theta)}_{\mu} = r(0,0,1,0), \nonumber\\
    && e^{(\varphi)}_{\mu} = r\sin\theta(-\omega,0,0,1), 
\end{eqnarray}
while the tetrad of the comoving frame of 1-forms takes then the form
\begin{eqnarray}
    && \bar{e}^{(t)}_{\mu}= e^{\Phi(r)/2}(1,0,0,\bar{\omega}),\nonumber\\
    && \bar{e}^{(r)}_{\mu}= e^{\Psi(r)/2}(0,1,0,0),\nonumber\\
    && \bar{e}^{(\theta)}_{\mu}= r(0,0,1,0),\nonumber\\
    && \bar{e}^{(\varphi)}_{\mu}= r\sin\theta(-\omega,0,0,1).
\end{eqnarray}

\subsection{Null geodesics of the first-order Hartle--Thorne--Tolman VII spacetimes}
The Lagrangian of the test particle (geodesic) motion in the internal and external first-order Hartle--Thorne spacetimes is given by \cite{cha:1983:BOOK} \footnote{We follow only procedure not the signature of metric tensor.}
\begin{equation}
2\pazocal{L}=g_{tt}\dot{t}^2+g_{rr}\dot{r}^2+g_{\theta\theta}\dot{\theta}^2+g_{\varphi\varphi}\dot{\varphi}^2+2g_{t\varphi}\dot{\varphi}\dot{t}.
\end{equation}
Metric functions Eq.~(\ref{metLTi}) and Eq.~(\ref{metLTe}) depend only on radial and latitudinal coordinates. Therefore, there exist two Killing vector fields (time and axial) that imply existence of two constants of motion
\begin{eqnarray}
&& p_t=\frac{\partial\pazocal{L}}{\partial\dot{t}}=g_{tt}\dot{t}+g_{t\varphi}\dot{\varphi}=-E=\mathrm{constant},\nonumber\\
&& p_{\varphi}=\frac{\partial\pazocal{L}}{\partial\dot{\varphi}}=g_{t\varphi}\dot{t}+g_{\varphi\varphi}\dot{\varphi}=\phi=\mathrm{constant}.
\label{e:constEL}
\end{eqnarray}
These are the energy $E$ and the axial component of the angular momentum $\phi$ of test particles (photons or neutrinos), as measured by static observers at infinity.

The test particle motion is governed by the Hamilton-Jacobi equation related to the metric \mbox{tensor $g^{\mu\nu}$}
\begin{equation}
2\frac{\partial S}{\partial \tau}=g^{\mu\nu}\frac{\partial S}{\partial x^{\mu}}\frac{\partial S}{\partial x^{\nu}},
\label{HJ}
\end{equation}
where $S$ denotes Hamilton's action function. The $g^{t\phi}$ metric coefficient is independent of latitude $\theta$, enabling thus separability of variables in the principal function  
\begin{equation}
S=\frac{1}{2}\delta\tau - Et + \phi\varphi + S_r(r) + S_{\theta}(\theta).
\label{action}
\end{equation}
The normalization condition is given in the form $p^{\mu}p_{\mu}=\delta$, where $\delta=-m^2$ for time-like geodesics, and $\delta=0$ for null geodesics. Introducing a separation constant $L$ serving as a motion constant related to the total angular momentum, all components of the particle four momentum can be expressed in the following way  
\begin{eqnarray}
&& p^r=g^{rr}p_r=g^{rr}\frac{\mathrm{d}S_r}{\mathrm{d}r}=\Big(2g^{t\varphi}g^{rr}E\phi+g^{rr}\delta-g^{tt}g^{rr}E^2-Lg^{rr}g^{\theta\theta} \Big)^{1/2},\label{e:pr}\\
&& p^{\theta}=g^{\theta\theta}p_{\theta}=g^{\theta\theta}\frac{\mathrm{d}S_{\theta}}{\mathrm{d}\theta}=g^{\theta\theta}\Bigg(L-\frac{\phi^2}{\sin^2\theta} \Bigg)^{1/2},\label{e:pth}\\
&& p^{\varphi}=\frac{\phi g_{tt}+Eg_{t\varphi}}{g_{tt}g_{\varphi\varphi}-g_{t\varphi}^2},\\
&& p^{t}=\frac{\phi g_{t\varphi}+Eg_{\varphi\varphi}}{g_{t\varphi}^2-g_{tt}g_{\varphi\varphi}}.
\end{eqnarray}
We focus attention on null geodesic. In the following, we consider the components of the four-momentum as components of wave-vector, $k^{\mu}$ and $k_{\mu}$. Further, the null geodesics (neutrino trajectories) are independent of the energy. We thus rescale the motion equations by energy and use new motion constants, namely the impact parameters 
\begin{eqnarray}
\lambda=\frac{\phi}{E}\quad \mathrm{and}\quad \pazocal{L}=\frac{L}{E^2}.
\label{e:constlal}
\end{eqnarray}
The motion in the radial and latitudinal directions ($r-\theta$) is restricted by the turning points, where $p^{r}=0$ or $p^{\theta}=0$. Using Eq.~(\ref{e:constlal}) in Eq.~(\ref{e:pr}( and Eq.~(\ref{e:pth}), we obtain the effective potentials governing the radial and latitudinal motion. For the radial motion we obtain general form
\begin{equation}
\pazocal{L}_{r}=2g^{t\varphi}g_{\theta\theta}\lambda-g^{tt}g_{\theta\theta},
\label{e:lr}
\end{equation}
and in the Hartle--Thorne--Tolman VII spacetimes we obtain the explicit form 
\begin{equation}
	\pazocal{L}_{r}(r,\lambda)=\cases{\frac{3r^2\left(1-2\omega\lambda\right)}{3-5\pazocal{C}}\cos^{-2}\left(C_a+Y(r)\right)\quad \textrm{for } r \le R\\
		\frac{r^3-4J\lambda}{r-2M} \quad \textrm{for } r>R.}
	\label{e:lr2}
\end{equation}
For the latitudinal motion, we arrive at the effective potential
\begin{equation}
\pazocal{L}_{\theta}(\theta,\lambda)=\frac{\lambda^2}{\sin^2\theta}.
\label{fcelth}
\end{equation}
The conditions $k^{r} \geq 0$ and $k^{\theta} \geq 0$ give restrictions on the impact parameter $\pazocal{L}$ in the form 
\begin{equation}
                 \pazocal{L}_r \geq \pazocal{L} \geq \pazocal{L}_{\theta}.
\end{equation}

The extreme points of the effective potential $\pazocal{L}_{r}(r)$ determine the circular null geodesics implying basic restrictions on the trapping effect. The local extrema of the effective potential $\pazocal{L}_{r}(r)$ (given by $d\pazocal{L}_{r}/dr=0$) are determined by the function $\lambda_c(r)$ determining the stable (internal at $r_\mathrm{c(s)}$) and unstable (external or internal at $r_\mathrm{c(u)}$) circular null geodesics. Using the relation Eq.~(\ref{e:lr2}), we arrive at general formula 
\begin{equation}
\lambda_{c}=\frac{\frac{\mathrm{d}}{\mathrm{d}r}(g^{tt}g_{\theta\theta})}{\frac{\mathrm{d}}{\mathrm{d}r}(2g^{t\varphi}g_{\theta\theta})}.
\label{lamc}
\end{equation}
In the external spacetime this formula corresponds to unstable circular null geodesics and takes the form 
\begin{equation}
\lambda_{c}(r,J)=-\frac{(r-3M) r^2}{2J} .  
\end{equation}
In the internal Tolman VII spacetime this formula can corresponds to both maxima (stable null geodesics) and minima (unstable null geodesics) of the effective potential,  having the form 
\begin{equation}
	\lambda_{c}(r;\pazocal{C},\omega)=\frac{h}{r \left(h_2+2 \sqrt{3}\right) \omega'(r)+2\omega(r) h},
\end{equation}
where
\begin{eqnarray}
	&&h_1(r;\pazocal{C})= \sqrt{3 \pazocal{C}^4 r^4-5 \pazocal{C}^2 r^2+\frac{1}{\pazocal{C}}},\nonumber \\
	&&h_2(r;\pazocal{C})= 6 \sqrt{3} \pazocal{C}^5 r^4-5 \pazocal{C} h_1+\pazocal{C}^3 r^2 \left(6 h_1-10 \sqrt{3}\right),\nonumber \\
	&&h(r;\pazocal{C})= h_2-\pazocal{C}^3 r^2 \left(6 \sqrt{3} \pazocal{C}^2 r^2+6 h_1-5 \sqrt{3}\right) \tan \Big(C_\mathrm{a} + Y(r) \Big)+2 \sqrt{3}.
\end{eqnarray}

The crucial restriction on the local extrema of the radial effective potential follows from the relation $\pazocal{L}_{r} \geq \pazocal{L}_{\theta}$. The corresponding limiting functions $\lambda_{r\pm}(r,\theta)$, following from the condition $\pazocal{L}_{r}=\pazocal{L}_{\theta}$, are given by the relation   
\begin{equation}
\lambda_{r\pm}= 
\sin^2{\theta}\Big( g^{t\varphi}g_{\theta\theta} \pm \sqrt{(g^{t\varphi}g_{\theta\theta})^2-g^{tt}g_{\theta\theta}\csc^2\theta} \Big)
\label{lamlim}
\end{equation}
that takes for the internal Tolman VII spacetime the form 
\begin{eqnarray}
	\lambda_{r\pm}^\mathrm{int}=&& \frac{3\omega(r)r^2\sin ^2\theta}{5\pazocal{C}-3} \cos^{-2} \Big(C_\mathrm{a} + Y(r) \Big)\nonumber\\
	&\pm& \sqrt{\frac{3\omega(r)}{3-5\pazocal{C}} }r\sin \theta\cos^{-1} \Big(C_\mathrm{a} + Y(r) \Big),
\end{eqnarray}
and for the external Lense-Thirring spacetime the form 
\begin{equation}
    \lambda_{r\pm}^\mathrm{ext}=\sin ^2\theta \left(-\frac{2 J}{r-2M}\pm \sqrt{\frac{r^3 \csc^2\theta}{r-2M}} \right).
\end{equation}
The radial effective function $\pazocal{L}_r$ is relevant in the regions restricted by the condition 
\begin{equation}
                 \lambda_{r-}(r,\theta) \leq \lambda \leq \lambda_{r+}(r,\theta). 
\end{equation}
								
\begin{figure}[ht]
	\centering
	\includegraphics[width=1.\hsize]{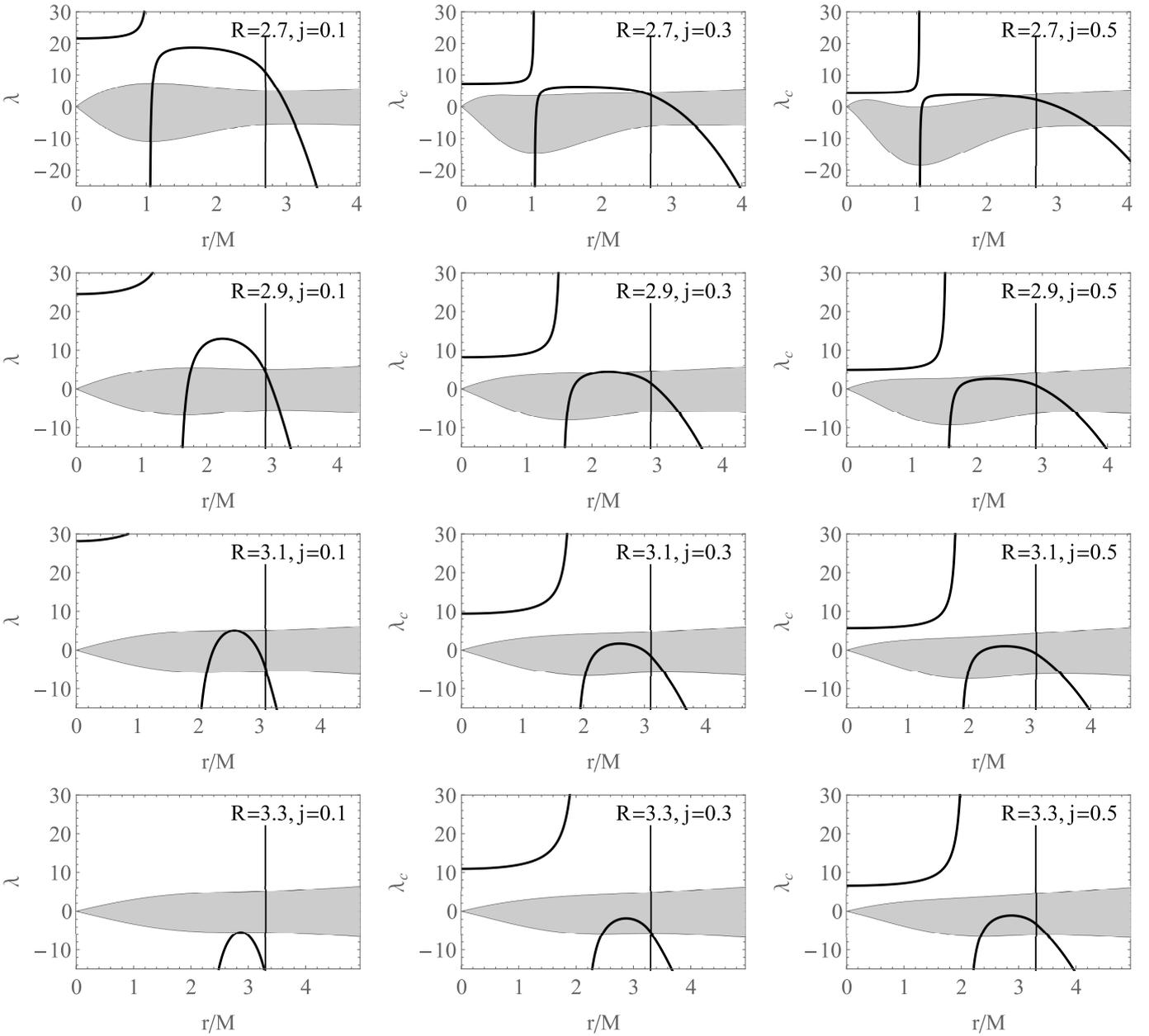}
	\caption{Restriction for the values of 
		$\lambda$ and the function for $\lambda_c$ in the equatorial 
		plane. The grey area indicates the allowed region for $\lambda$ 
		with the boundaries given by $\lambda_{r+}$ and $\lambda_{r-}$. The 
		bold solid line shows the values of $\lambda_c$ which marks the 
		extremes of $\pazocal{L}$. The first column is for $j=0.1$, the 
		second is for $j=0.3$ and the last is for $j=0.5$. The first row 
		is for $R/M=2.7$, the second is for $R/M=2.9$, the third is for 
		$R/M=3.1$ and the last is for $R/M=3.3$.}
	\label{f:lclthj}
\end{figure}

In the following we restrict discussion to the most extended region in the equatorial plane with $\sin\theta=1$, where the trapping effects are most profound. In Fig.~\ref{f:lclthj} we illustrate both the functions $\lambda_{r\pm}(r,\theta=\pi/2)$ and $\lambda_{c}(r)$ for characteristic values of the spacetime parameters, namely for different rotation rates $j$ ($j=J/M^2$ is the dimensionless angular momentum) and inverse compactness of the object $R/M$. The motion with a fixed value of the impact parameter $\lambda$ is allowed in the regions located between the curves given by Eq.~(\ref{lamlim}) (grey area). The extrema points of the radial effective potential, corresponding to the circular null geodesics, are given by the function $\lambda_{c}(r)$. 

The stable circular null geodesic for highly compact object with $R/M=2.6$ is located in an almost constant position for most allowed $\lambda_c$, but the unstable circular null geodesic changes its position significantly within the allowed $\lambda_c$. We can also point out the non-existence of circular null geodesics for low rotations and slightly compact objects with $R/M=3.2$ -- as expected because for non-rotating configurations with $R/M>3.202$, the circular null geodesics do not exist. 

\subsection{Effective potential of the radial motion}

The trapping effect is governed by the effective potential of the radial motion $\pazocal{L}_{r}(r,\lambda)$ given by Eq.~(\ref{e:lr}), with its dependence on the impact parameter $\lambda$ of the null geodesic; and the dependence on the latitudinal coordinate reflecting restrictions on the allowed range of values of $\lambda$. 
As the influence of rotation is vanishing ($g^{t\varphi}=0$) at the poles, Eq.~(\ref{e:lr}) matches Eq.~(\ref{e:sveff}), the effective potential reduces to those related to the internal Tolman VII spacetime, and the external Schwarzschild spacetime. Therefore, our results at the poles have to coincide with those obtained in the spherically symmetric Tolman VII spacetimes. 

\begin{figure}[ht]
	\centering
	\includegraphics[width=0.6\hsize]{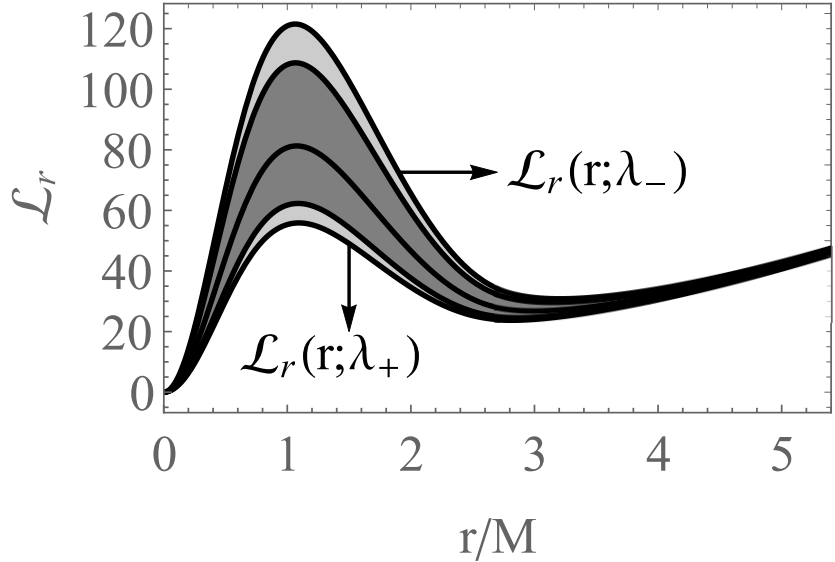}
	\caption{The effective potential
		$\pazocal{L}_r$ for allowed values of the
		parameter $\lambda$. In the shaded area, we find all possible
		effective potentials for the allowed values of lambda and
		$\theta = \pi / 2$. In the darker region, there are effective
		potentials with allowed values $\lambda$ and $\theta = \pi / 4$
		or $3 \pi / 4$, and the curve in the middle of the dark region
		plots the effective potential as $\theta$ approaches
		the poles.}
	\label{f:Vefflim}
\end{figure}

In the first-order Hartle--Thorne--Tolman VII spacetime the effective potential depends on the parameter $\lambda$ determining the motion in the axial direction. At a given radius $r$ and latitude $\theta=\pi/2$ corresponding to the equatorial plane of the rotating object, the limiting values of $\lambda$ given by the functions $\lambda_{r\pm}(r,\theta=\pi/2)$ govern the maximally co-rotating ($\lambda_{r+}(r,\theta=\pi/2)$) and maximally counter-rotating ($\lambda_{r-}(r,\theta=\pi/2)$) null geodesics. The effective potentials for such impact parameters, and the related directions, represent the limiting profile for the effective potentials corresponding to the null geodesics in the other directions with impact parameters $\lambda \in (\lambda_{r-}(r,\theta=\pi/2),\lambda_{r-}(r,\theta=\pi/2))$. In the case of $\theta \neq \pi/2$, the limiting impact parameters $\lambda_{r\pm}(r,\theta)$ are determined by Eq.~(\ref{lamlim}), where the shifting factor $\sin^2\theta$ occurs being related to the vanishing of the rotational effects on the rotation axis $\theta=0$ -- see Fig.~\ref{f:Vefflim}. 

The location of the effective potential $\pazocal{L}_{r}(r,\lambda)$ maximum and minimum, corresponding to the locations $r_\mathrm{c(s)}$and $r_\mathrm{c(u)}$ of the circular null geodesics, can be easily obtained from the functions $\lambda_{c}(r)$ and from Fig.~\ref{f:lclthj}. In the first order Hartle--Thorne--Tolman VII spacetimes the maximum is always below the surface of the object, while the minimum can be both above and below the surface. These extrema depend not only on the parameters of the spacetime, but also on the impact parameter $\lambda$ (see Fig.~\ref{f:Vefflim}), similarly to the radii giving the limits on the trapping region -- the radius governing motion fully constrained to the internal spacetime, $r_\mathrm{c(s)}$, that is given by equation $\pazocal{L}_{r}(r,\lambda) = \pazocal{L}_{r}(R,\lambda)$, and the radius allowing for motion in the external spacetime (if relevant), $r_\mathrm{c(u)}$, that is solution of $\pazocal{L}_{r}(r,\lambda) = \pazocal{L}_r(r_\mathrm{c(u)},\lambda)$. These details have to be taken into account in calculation of coefficients characterizing the trapping effect globally, for complete first-order Hartle--Thorne--Tolman VII spacetimes. 

\begin{figure}[ht]
	\centering
	\includegraphics[width=0.8\hsize]{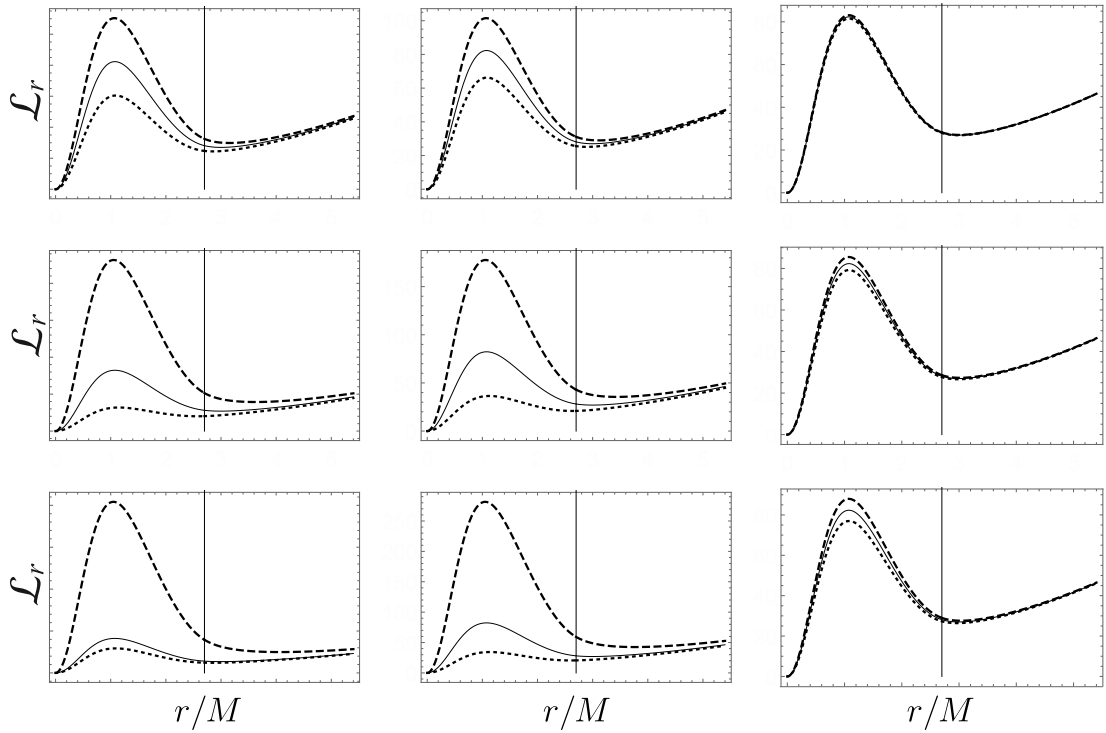}
	\caption{Effective potentials of the first-order Hartle--Thorne--Tolman VII objects
		with inverse compactness $R/M=2.7$ given for several values of $j$ and
		$\theta$. The solid lines are effective
		potentials for non-rotating configurations.
		The dotted and dashed lines denote
		effective potentials for co-rotating and
		counter-rotating null geodesics 
		($\lambda=\lambda_{r+}$ and
		$\lambda=\lambda_{r-}$ respectively). The first column is for
		$\theta=\pi/2$, the second is for $\theta=\pi/4$ and
		the last is for $\theta=1/100$. The
		first row is for $j=0.1$, the second is for
		$j=0.3$ and the last is for $j=0.5$.}
	\label{f:rm27}
\end{figure}

\begin{figure}[ht]
	\centering
	\includegraphics[width=0.8\hsize]{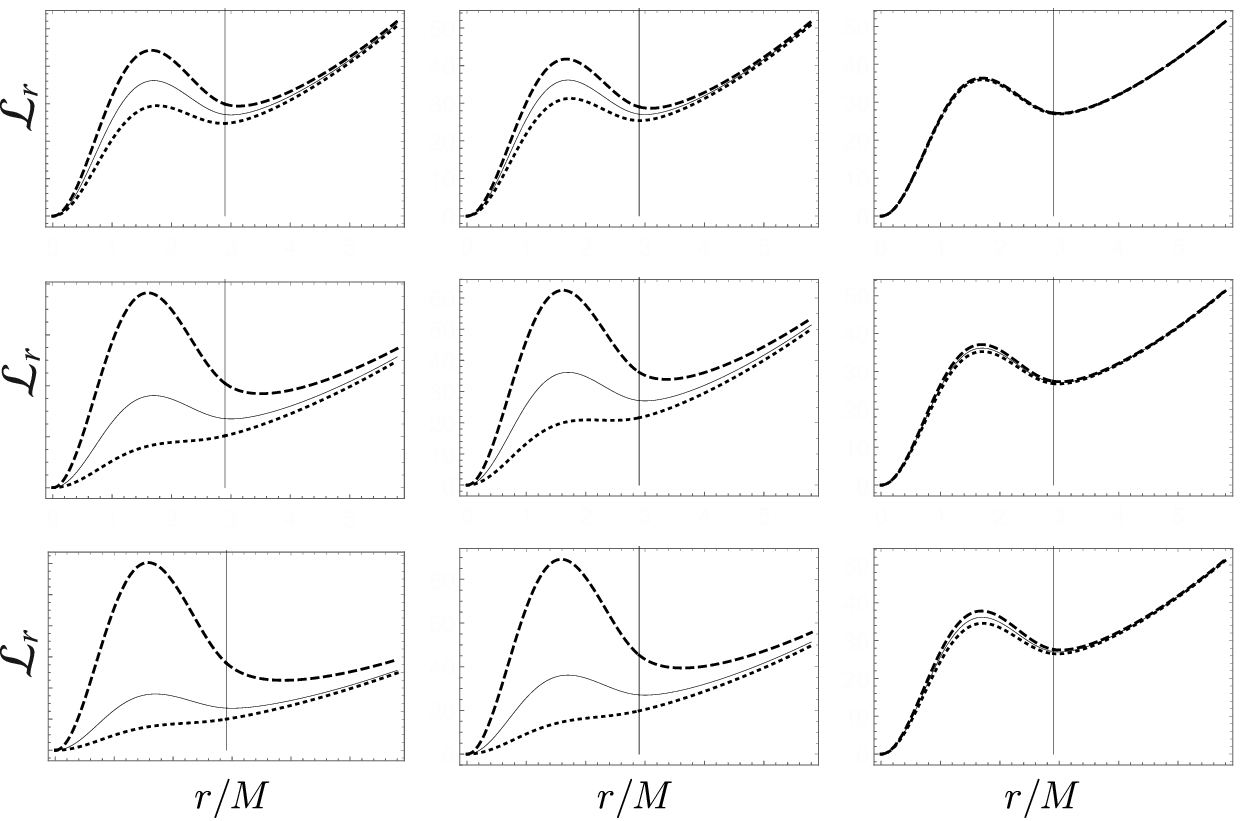}
	\caption{Effective potentials of the first-order Hartle--Thorne--Tolman VII objects
		with inverse compactness $R/M=2.9$ given for several values of $j$ and
		$\theta$. The choice of $j$ and $\theta$ is the same as in Fig.~\ref{f:rm27}.}
	\label{f:rm29}
\end{figure}

\begin{figure}[ht]
	\centering
	\includegraphics[width=0.8\hsize]{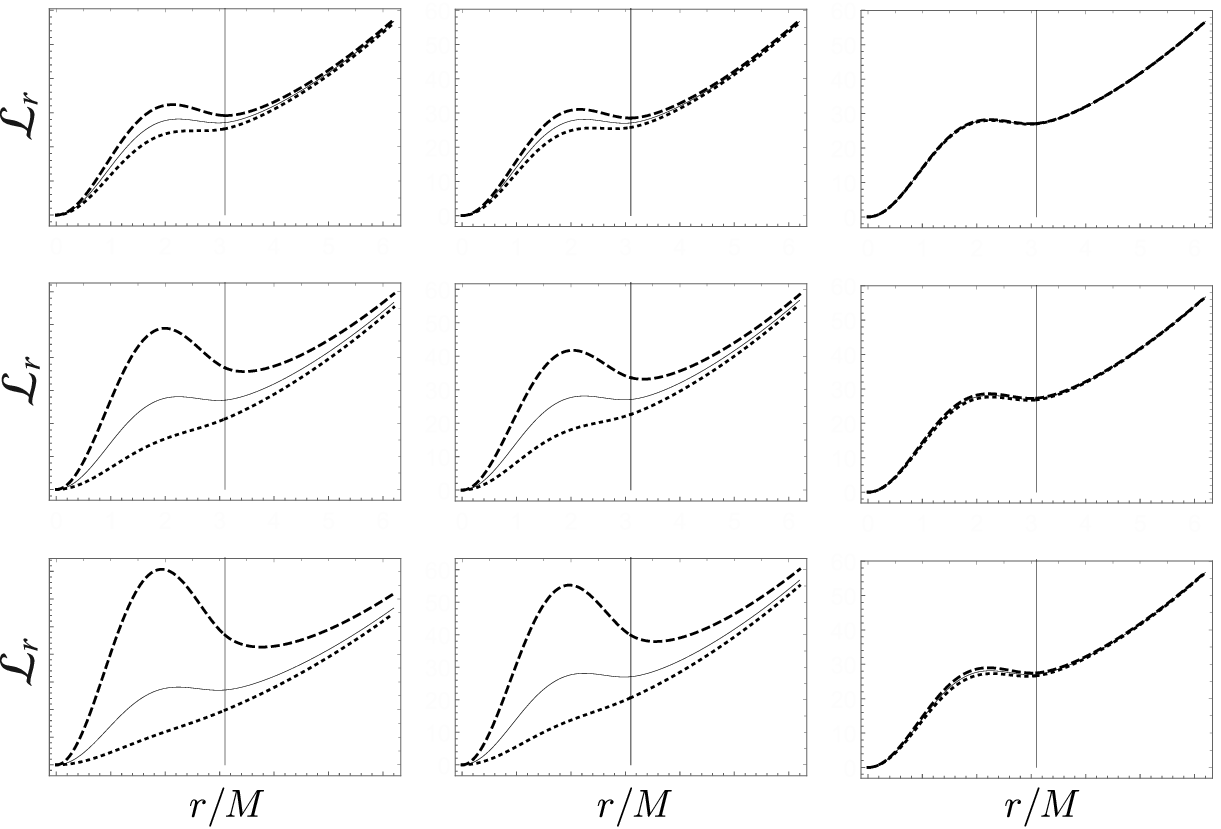}
	\caption{Effective potentials of the first-order Hartle--Thorne--Tolman VII objects
		with inverse compactness $R/M=3.1$ given for several values of $j$ and
		$\theta$. The choice of $j$ and $\theta$ is the same as in Fig.~\ref{f:rm27}.}
	\label{f:rm31}
\end{figure}

\begin{figure}[ht]
	\centering
	\includegraphics[width=0.8\hsize]{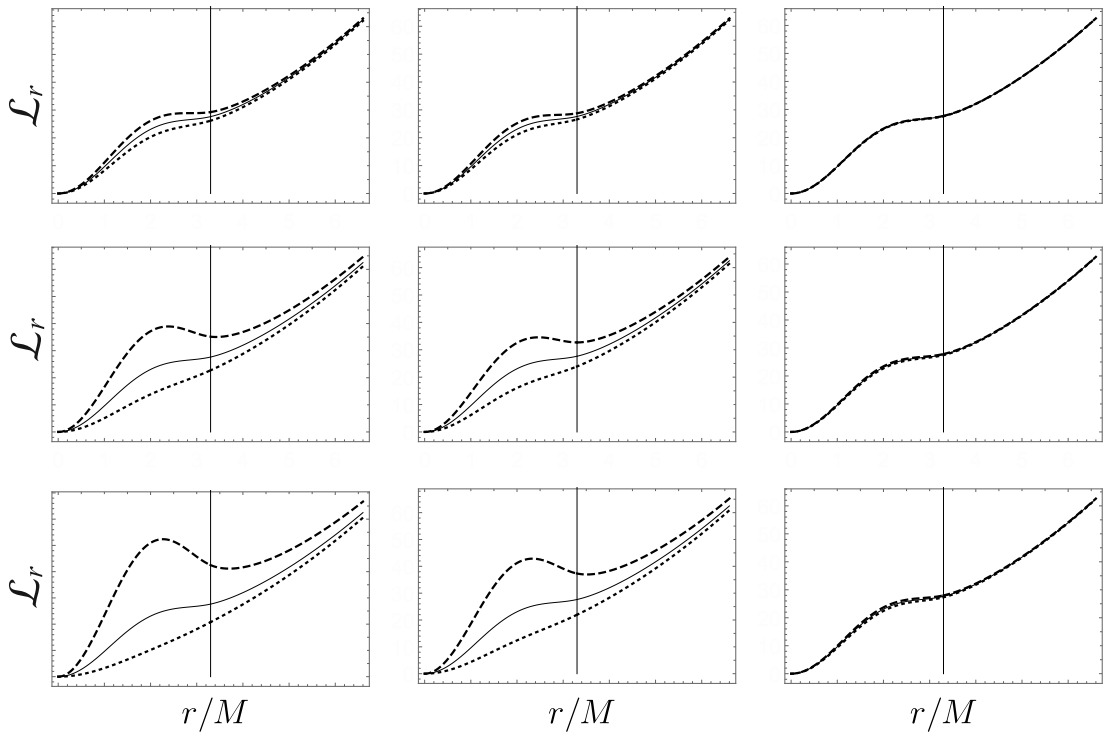}
	\caption{Effective potentials of the first-order Hartle--Thorne--Tolman VII objects
		with inverse compactness $R/M=3.3$ given for several values of $j$ and
		$\theta$. The choice of $j$ and $\theta$ is the same as in Fig.~\ref{f:rm27}.}
	\label{f:rm33}
\end{figure}

The effective potentials corresponding to the maximally co-rotating and maximally counter-rotating null geodesics (the limiting values of the impact parameter $\lambda$) are given for various rotation rates represented by the parameter $j$, the spacetime inverse compactness $R/M$, and characteristic latitudes $\theta$ in figures \ref{f:rm27} - \ref{f:rm33}; their comparison with the non-rotating effective potential corresponding to the spaherically symmetric Tolman VII spacetimes is also presented -- as expected, the effective potential near the poles approaches the effective potential related to the non-rotating Tolman VII spacetime as the rotation does not manifest itself at poles ($g_{t\varphi}(r,\theta=0)=0$). To stress and clearly illustrate the role of the rotation in the trapping effects, we consider also relatively high values of the rotational parameter $j$, demonstrating how increasing rotation causes increase of trapping of counter-rotating null geodesics, while it decreases trapping of co-rotating null geodesics. This phenomenon is enhanced with decreasing $R/M$. 

The monotonicity of the effective potential in the Tolman VII spacetimes with $R/M>3.202$ results in vanishing of the trapping effects in rotating configurations with $R/M>3.202$ near the poles, independently of the rotation rate. In the configurations with $R/M>3.202$ the trapping effect arise for the counter-rotating null geodesics, but vanish for the co-rotating null geodesics, and efficiency of the trapping increases with increasing rotation rate $j$. 

\FloatBarrier

\begin{figure}[ht]
	\centering
	\includegraphics[width=1.\hsize]{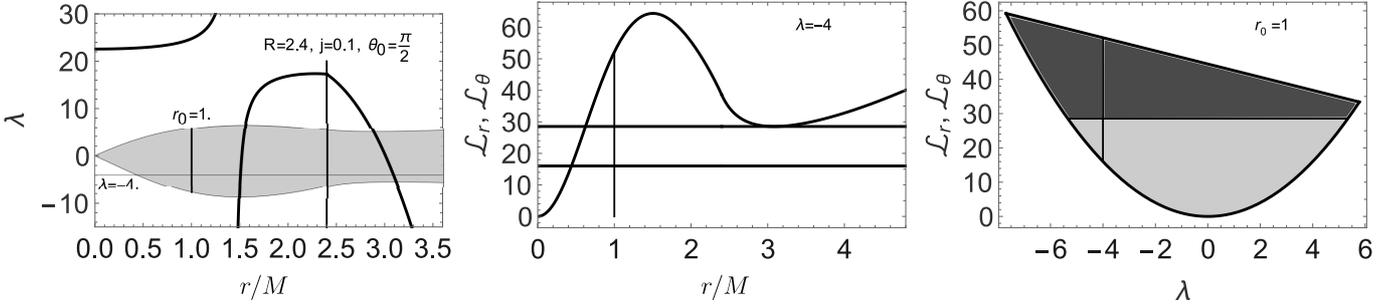}
	\caption{The left panel determines restriction on values of $\lambda$ and gives the function $\lambda_c$. The middle panel shows allowed values of $\pazocal{L}$ in dependence on $\lambda$. The right panel depict the effective potentials of the radial and latitudinal motion. All figures describes specific case of emitting null geodesics from equatorial plane and radius $r_0=1$ with $\lambda=-4$ which corresponds to the minimum of the effective potential $\pazocal{L}_r$.}
	\label{f:Lllc}
\end{figure}

The process of construction of the trapping cones (and complementary escape cones) is illustrated in Fig.~\ref{f:Lllc} demonstrating the characteristic functions of $\lambda$ give the corresponding effective potentials $\pazocal{L}_r$ that enables to determine in a given position characterized by coordinates $r,\theta$ the range of values of the impact parameters $\lambda$ and $\pazocal{L}$ related to the trapped (escaping) null geodesics.

\section{Trapping cones in the first-order Hartle--Thorne--Tolman VII spacetimes}

In the spherically symmetric Tolman VII spacetimes the trapping (escape) cones are centrally symmetric as they are dependent on a single impact parameter due to the symmetry, but in the axially symmetric Hartle--Thorne--Tolman VII internal spacetimes the symmetry is naturally broken, so they are governed by two impact parameters (or two directional angles). 
 
The null geodesics are fully determined by the two impact parameters $\lambda$ and $\pazocal{L}$ that have to be related to any pair of the directional angles \{$\alpha,~\beta,~\gamma$\}. The tetrad components of the wave four-vector are related to the directional angles in the tetrad due to the relations 
\begin{eqnarray}
&&k^{(t)}=-k_{(t)}=1,\\
&&k^{(r)}=k_{(r)}=\cos{\alpha},\\
&&k^{(\theta)}=k_{(\theta)}=\sin{\alpha}\cos{\beta},\\
&&k^{(\varphi)}=k_{(\varphi)}=\sin{\alpha}\sin{\beta}=\cos{\gamma}.
\label{e:mom}
\end{eqnarray}
The directional angle $\alpha$ is given by 
\begin{eqnarray}
\cos{\alpha}=\frac{k^{(r)}}{k^{(t)}}=\frac{\bar{e}^{(r)}_{\mu}k^{\mu}}{\bar{e}^{(t)}_{\mu}k^{\mu}} . 
\label{e:cosa}
\end{eqnarray}
Substituting Eq.~(\ref{e:mom}) in Eq.~(\ref{e:constlal}) we can express the impact parameter $\lambda$ in terms of the angle $\gamma$
\begin{eqnarray}
\lambda=\frac{\bar{e}_{\varphi}^{(\mu)}k_{(\mu)}}{-\bar{e}_{t}^{(\mu)}k_{(\mu)}}=\frac{\bar{e}_{\varphi}^{(t)}k_{(t)}+\bar{e}_{\varphi}^{(\varphi)}k_{(\varphi)}}{-\bar{e}_{t}^{(t)}k_{(t)}-\bar{e}_{t}^{(\varphi)}k_{(\varphi)}}=\frac{-\bar{e}_{\varphi}^{(t)}+\bar{e}_{\varphi}^{(\varphi)}\cos{\gamma}}{\bar{e}_{t}^{(t)}-\bar{e}_{t}^{(\varphi)}\cos{\gamma}}.
\label{e:lamgam}
\end{eqnarray}

Construction of the trapping (escaping) is done in close analogy with the method developed and used in \cite{Sch-Stu:2009:IJMPD,Stu-Sch:2010:CQG,Stu-Cha-Sch:2018:EPJC}. At a fixed point of the first-order Hartle--Thorne--Tolman VII spacetime, characterized by coordinates ($r,\theta$), we set the angle $\gamma$ and determine the impact parameter $\lambda$ and the related effective potential $\pazocal{L}_{r}(r,\lambda)$ governing the restrictions on the impact parameter $\pazocal{L}$. To construct the trapping (escape) cone, we have to know the minimum of $\pazocal{L}_r$ given by$\pazocal{L}_r/$d$r$=0. The intersection of $\pazocal{L}(r_\mathrm{c(u)},\lambda)$ ($\pazocal{L}_r$ in minimum) and $\pazocal{L}_\theta(\lambda)$ (which gives the minimal allowed value of $\pazocal{L}$) implies a restriction on the values of $\lambda$ and due to Eq.~(\ref{e:lamgam}) we arrive to the restriction on the directional angle $\gamma$. (The allowed $\lambda$ values are presented by the gray area in the left panel of Fig.~\ref{f:Lllc}.) 

The directional angles $\gamma\in \langle\gamma_{min},\gamma_{max}\rangle$ represent the cut-off values separating null geodesics escaping to infinity and those trapped by gravity. From Eq.~(\ref{e:cosa}) we obtain the limiting value of the angle $\alpha$, and from Eq.~(\ref{e:mom}) we obtain also the limiting value of the angle $\beta$. The boundary of the trapping cone is formed by plotting all the limiting angles [$\alpha, \beta$]. The region of co-rotating (counter-rotating) null geodesics contains the angle $\beta=\pi/2$ ($\beta=3\pi/2$), the separation line is given by angles $\beta=0$ and $\beta=\pi$.  

\begin{figure}[ht]
	\centering
	\includegraphics[width=0.90\hsize]{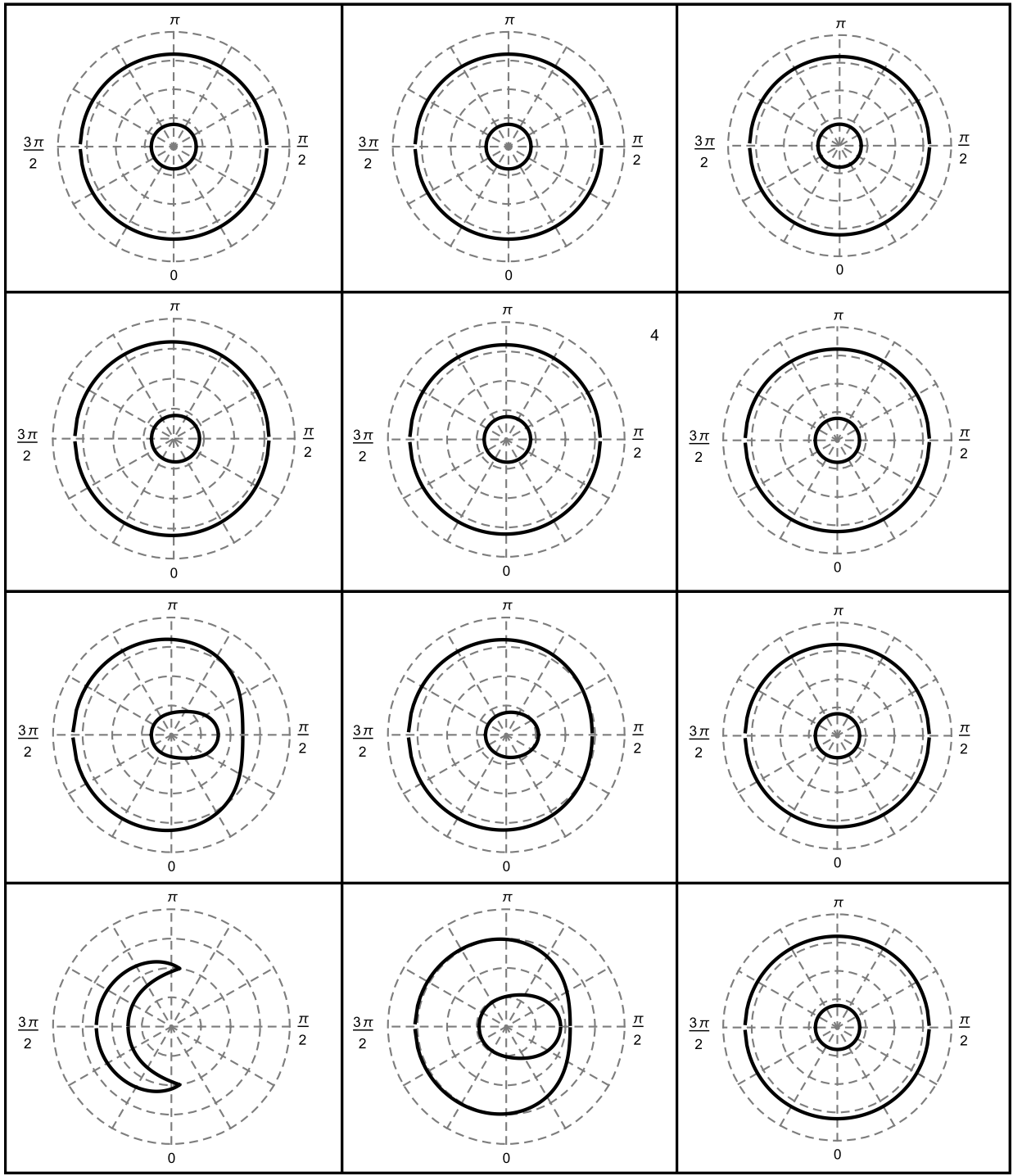}
	\caption{Escape cones of the Hartle--Thorne--Tolman VII objects with 
		compactness $R/M=2.7$ given for 
		several values of $\theta$ and $j$. The first column is for 
		$\theta=\pi/2$, the second column is for $\theta=\pi/4$ and 
		the last column is for $\theta=1/100$. The first row is for $j=0.01$, 
		the second is for $j=0.1$, the third is for $j=0.3$ and the 
		last is for $j=0.5$.}
	\label{krm27}
\end{figure}

\begin{figure}[ht]
	\centering
	\includegraphics[width=0.90\hsize]{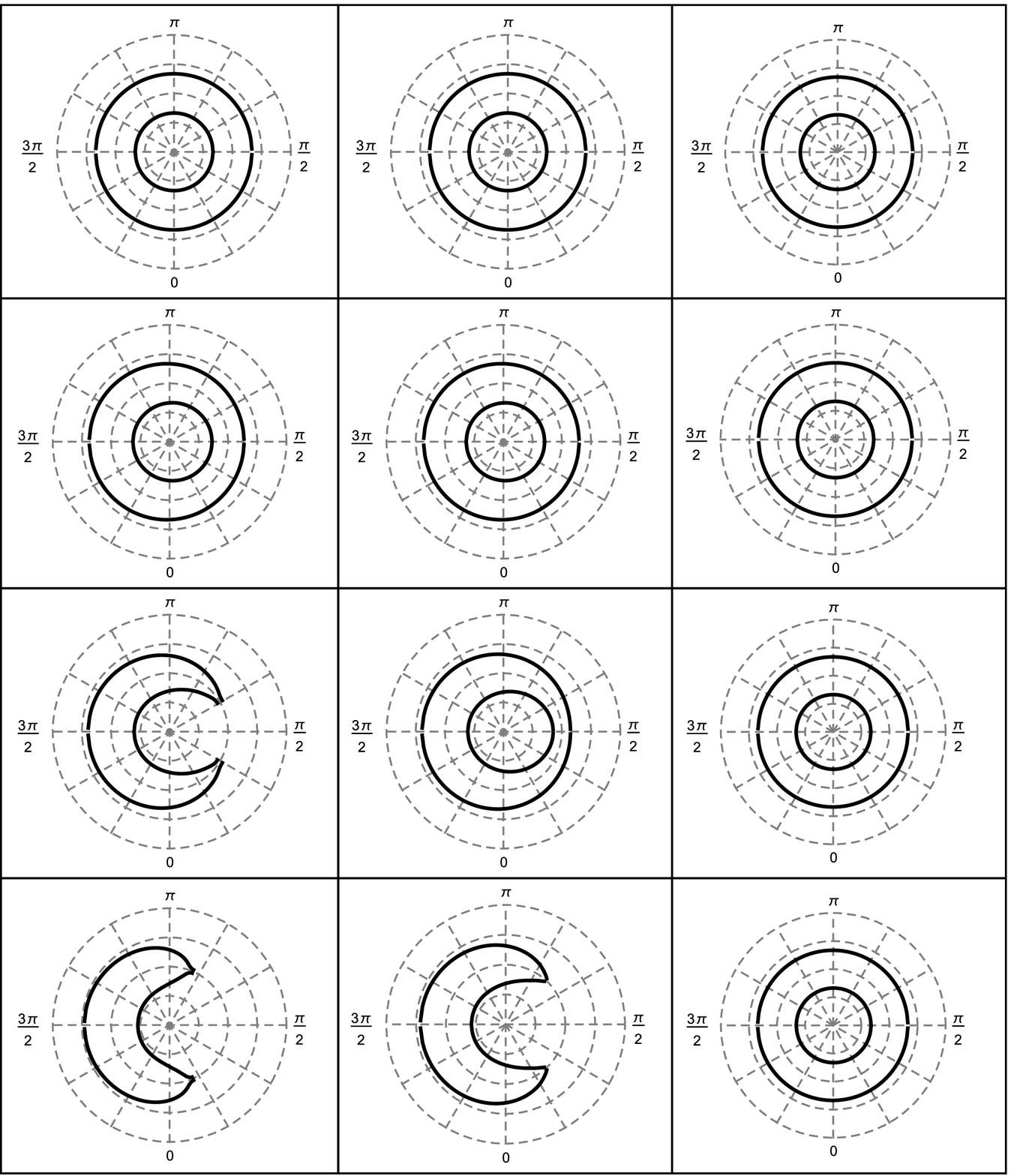}
	\caption{Escape cones of the Hartle--Thorne--Tolman VII objects with 
		compactness $R/M=2.9$ given for 
		several values of $j$ and $\theta$. The choice of $j$ and $\theta$ is the same as in Fig.~\ref{krm27}.}
	\label{krm29}
\end{figure}

\begin{figure}[ht]
	\centering
	\includegraphics[width=0.90\hsize]{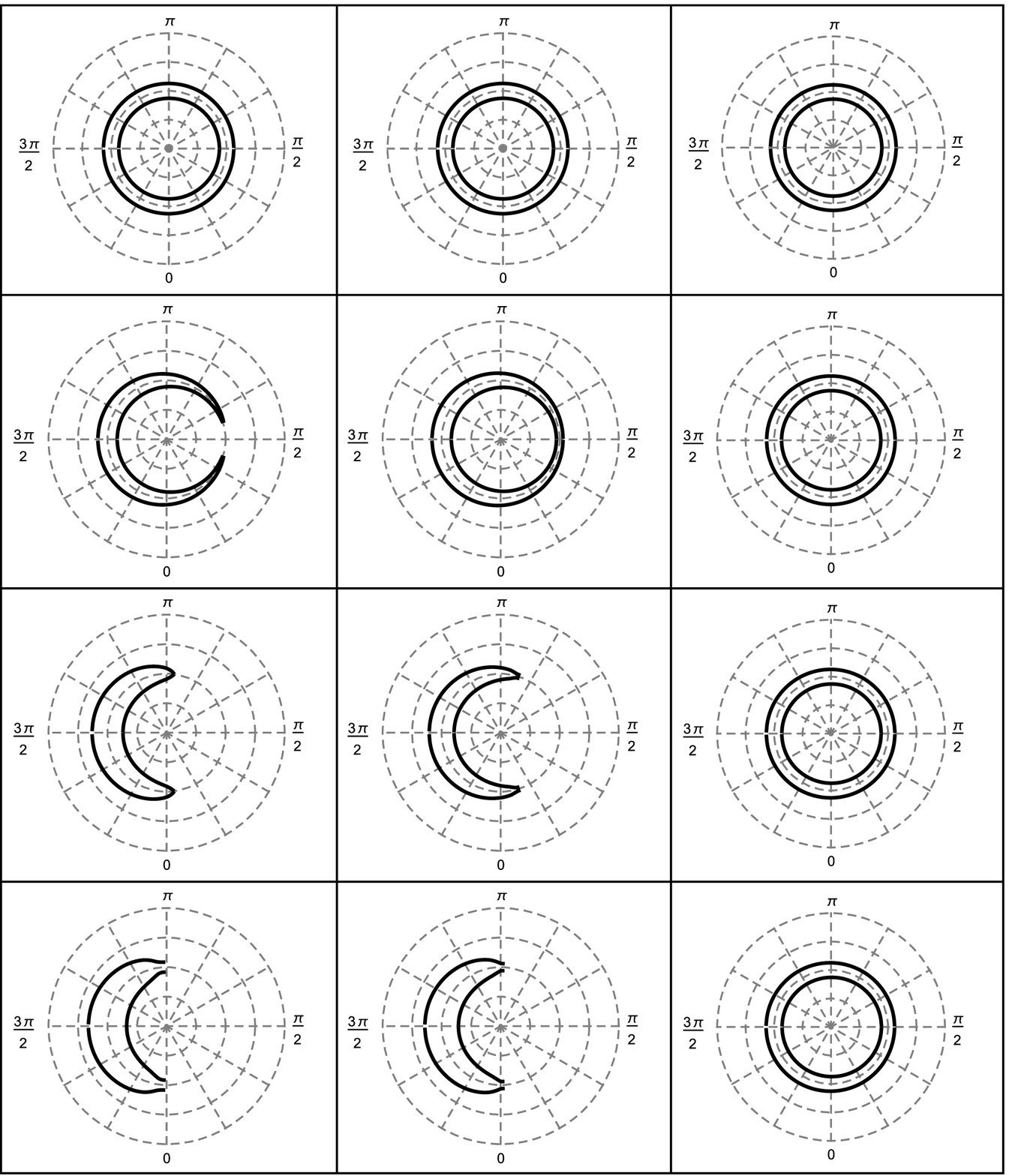}
	\caption{Escape cones of the Hartle--Thorne--Tolman VII objects with 
		compactness $R/M=3.1$ given for 
		several values of $j$ and $\theta$. The choice of $j$ and $\theta$ is the same as in Fig.~\ref{krm27}.}
	\label{krm31}
\end{figure}

\begin{figure}[ht]
	\centering
	\includegraphics[width=0.90\hsize]{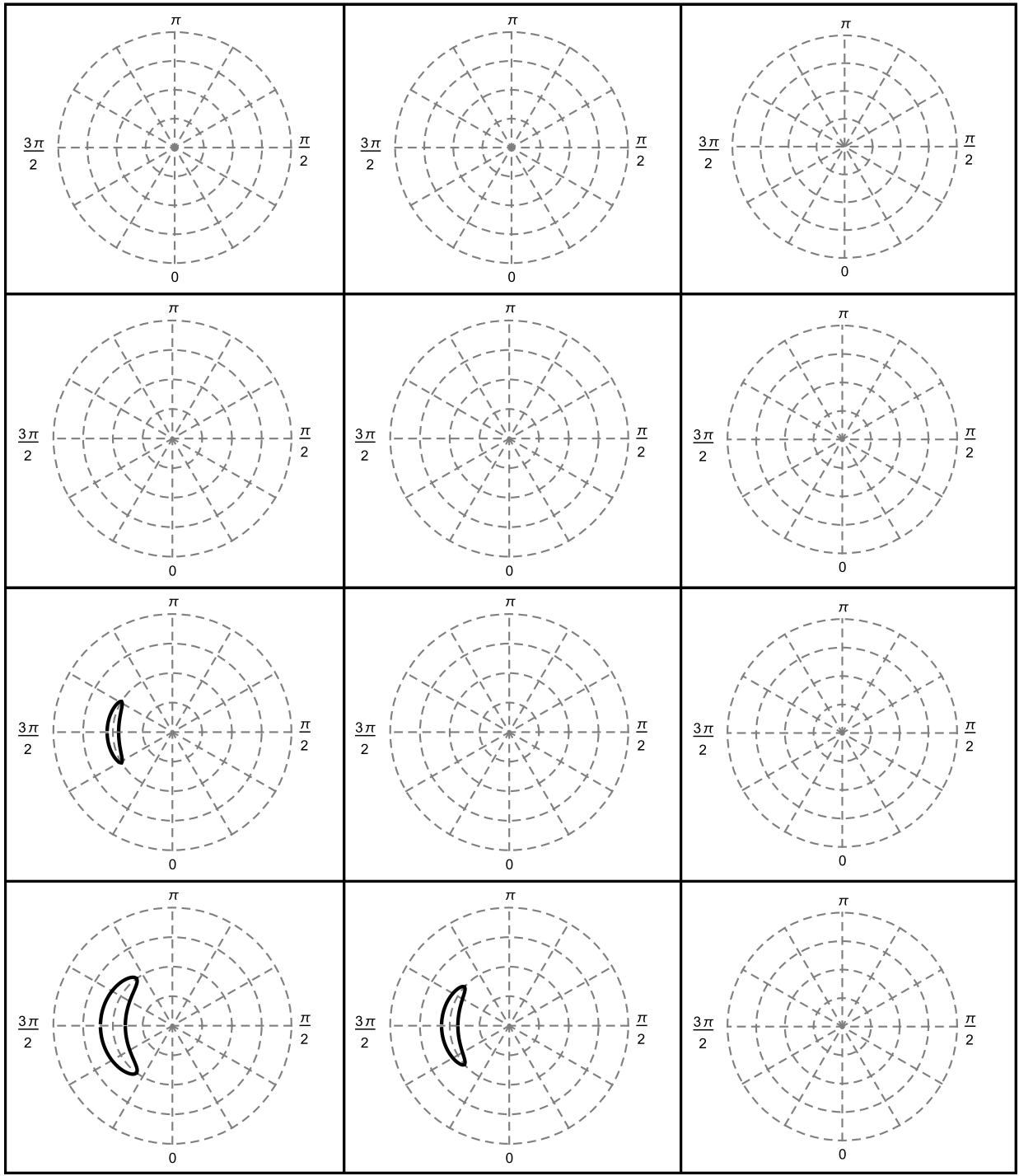}
	\caption{Escape cones of the Hartle--Thorne--Tolman VII objects with 
		compactness $R/M=3.3$ given for 
		several values of $j$ and $\theta$. The choice of $j$ and $\theta$ is the same as in Fig.~\ref{krm27}.}
	\label{krm33}
\end{figure}

The constructed cones are presented in Figs.\ref{krm27} - \ref{krm33} for the same selection of the spacetime parameters and the latitudes of the cone position as in the case of the representative figures of the effective potentials; the radial coordinate is fixed at $r/M=1$. These figures clearly demonstrate the breaking of the central symmetry, increasing with increasing values of $R/M$ and decreasing with decreasing value of the latitudinal coordinate of the position of the trapping cone, as the rotation effects are weakening as the point where the cone is constructed approaches the symmetry axis. 

The monotonic effective potential for the spacetimes with small rotation rate and $R/M>3.202$ means disappearance of the trapping cone that can exist only in small region of counter-rotating null geodesics in spacetimes with sufficiently high rotation parameter $j$. 

\FloatBarrier

\section{Efficiency of trapping}

The trapping of neutrinos in the first-order Hartle--Thorne--Tolman VII spacetimes can be characterized by the local and global trapping efficiency coefficients that can be defined in the form introduced in \cite{Vrb-Urb-Stu:2020:EPJC}, slightly modified in comparison with those introduced in the spherically symmetric trapping spacetimes \cite{Stu-Tor-Hle:2009:CQG,Stu-Vrb-Hla:2021:sub1}. We assume locally isotropic sources of the null geodesics representing radiated neutrinos, as in \cite{Stu-Tor-Hle:2009:CQG,Stu-Vrb-Hla:2021:sub1}.
 
\subsection{Local trapping coefficient}

This coefficient gives in the region allowing for the trapping phenomena the radial profiles of the local trapping efficiency, at fixed latitudinal coordinates of the compact object; we use the approach introduced in \cite{Stu-Tor-Hle:2009:CQG}, with the assumption of the isotropically radiating sources.  

To properly illustrate the local trapping coefficient, we variate $\theta$, $j$ and $R/M$ as in the representative cases of the effective potential of the null geodesic motion $\pazocal{L}_{r}(r,\lambda)$ and the trapping (escape) cones. 

\begin{figure}[ht]
	\centering
	\includegraphics[width=0.90\hsize]{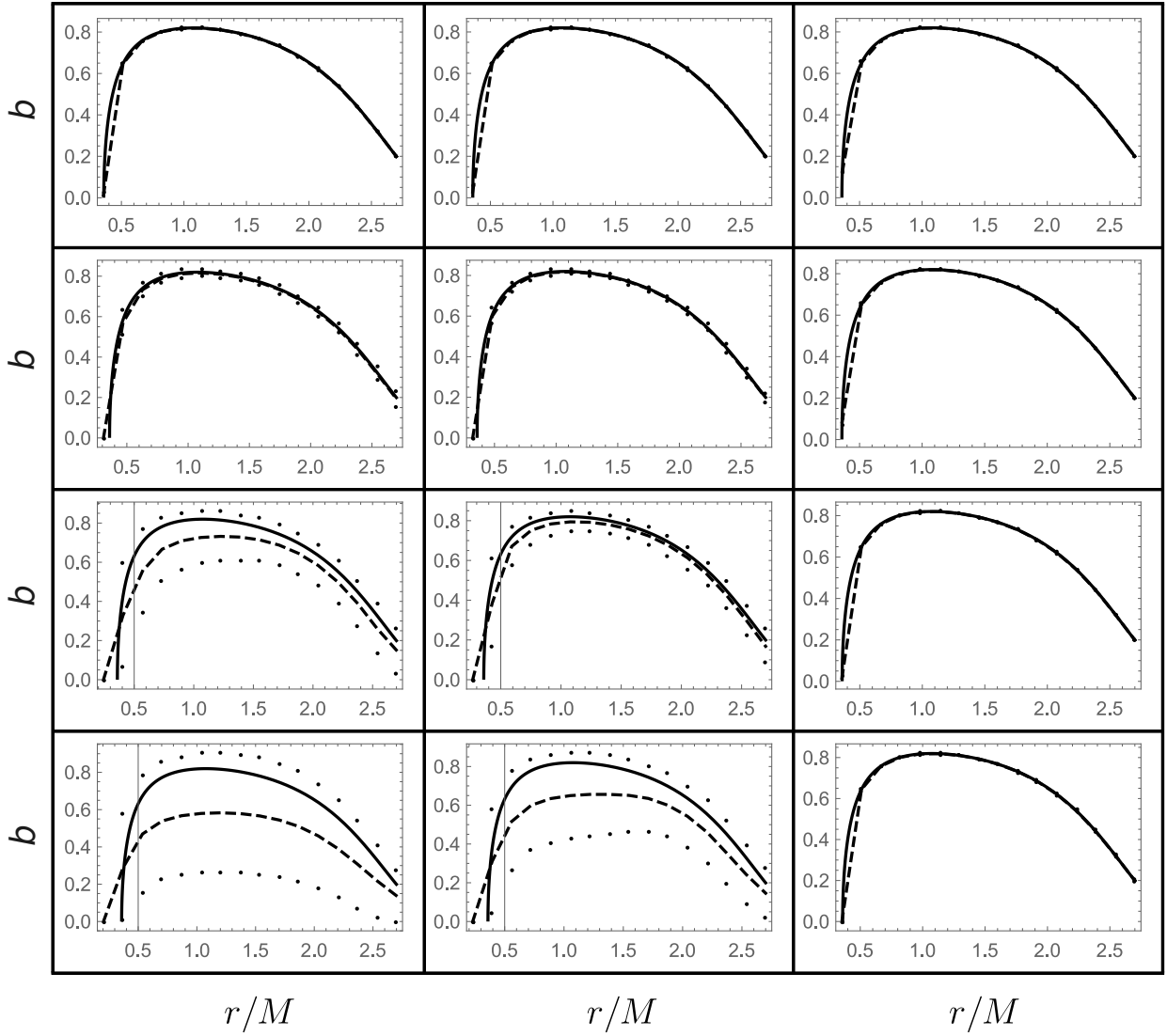}
	\caption{The radial profile of the local trapping efficiency coefficient $b$ is given for for the first-order Hartle--Thorne--Tolman VII object with 
		$R/M=2.7$ is plotted for several values of $j$ and $\theta$. 
		The solid line shows the local trapping for a non-rotating 
		configuration and the dashed line shows local trapping 
		for the rotating case. For the latter, 
		the points above the dashed line are values for just 
		counter-rotating directions of the null 
		geodesics, and the points below it are those for 
		co-rotating directions. The first column 
		is for $\theta=\pi/2$, the second is for $\theta=\pi/4$ and 
		the last is for $\theta=1/100$. The first row is for $j=0.01$, 
		the second is for $j=0.1$, the third is for $j=0.3$ and the 
		last is for $j=0.5$.}
	\label{loct27}
\end{figure}

\begin{figure}[ht]
	\centering
	\includegraphics[width=0.90\hsize]{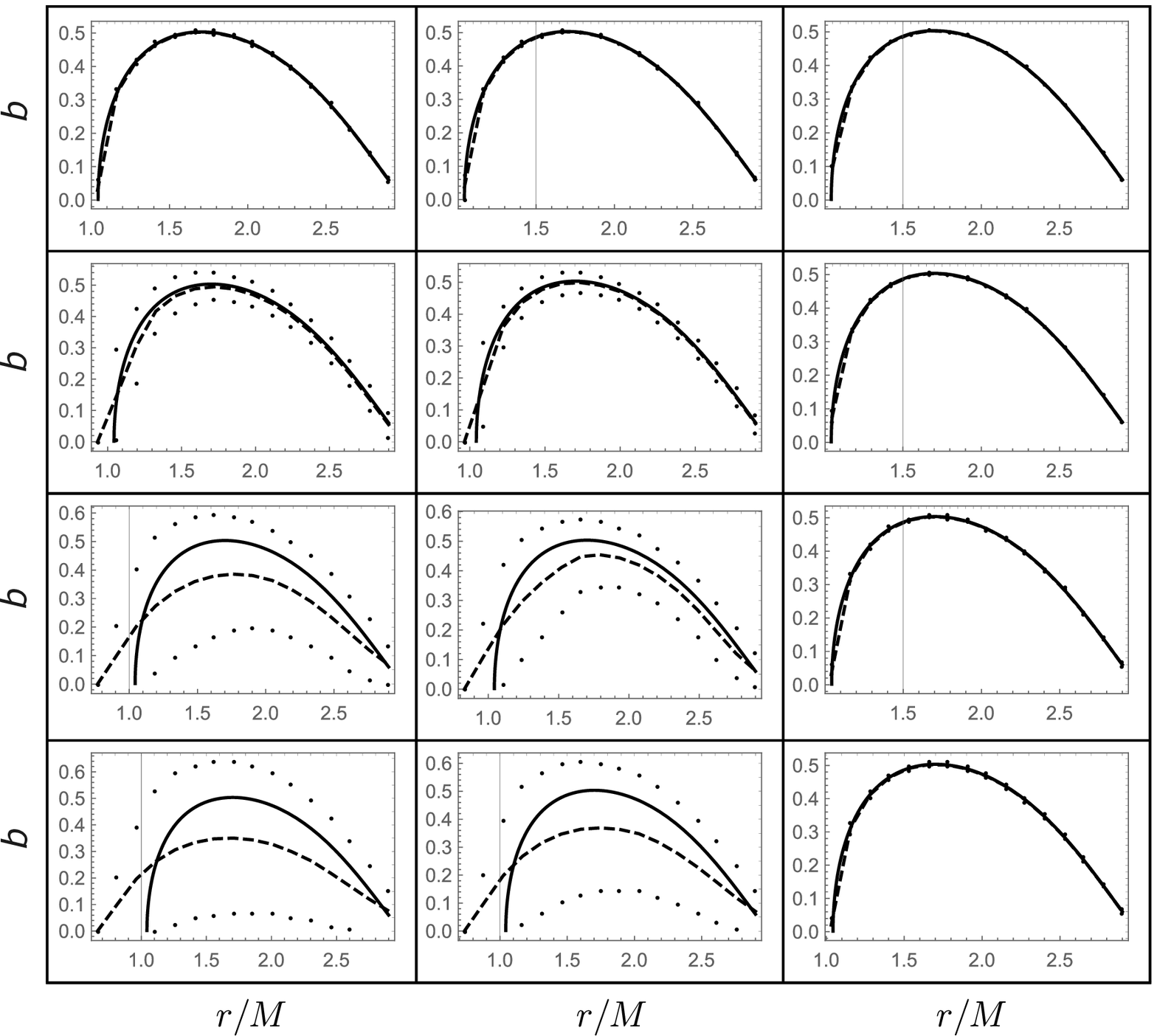}
	\caption{The radial profile of the local trapping efficiency coefficient $b$ is given for for the first-order Hartle--Thorne--Tolman VII object with 
		$R/M=2.9$ is plotted for several values of $j$ and $\theta$. 
		The curves and a choice of $j$ and $\theta$ are the same as in Fig.~\ref{loct27}.}
	\label{loct29}
\end{figure}

\begin{figure}[ht]
	\centering
	\includegraphics[width=0.90\hsize]{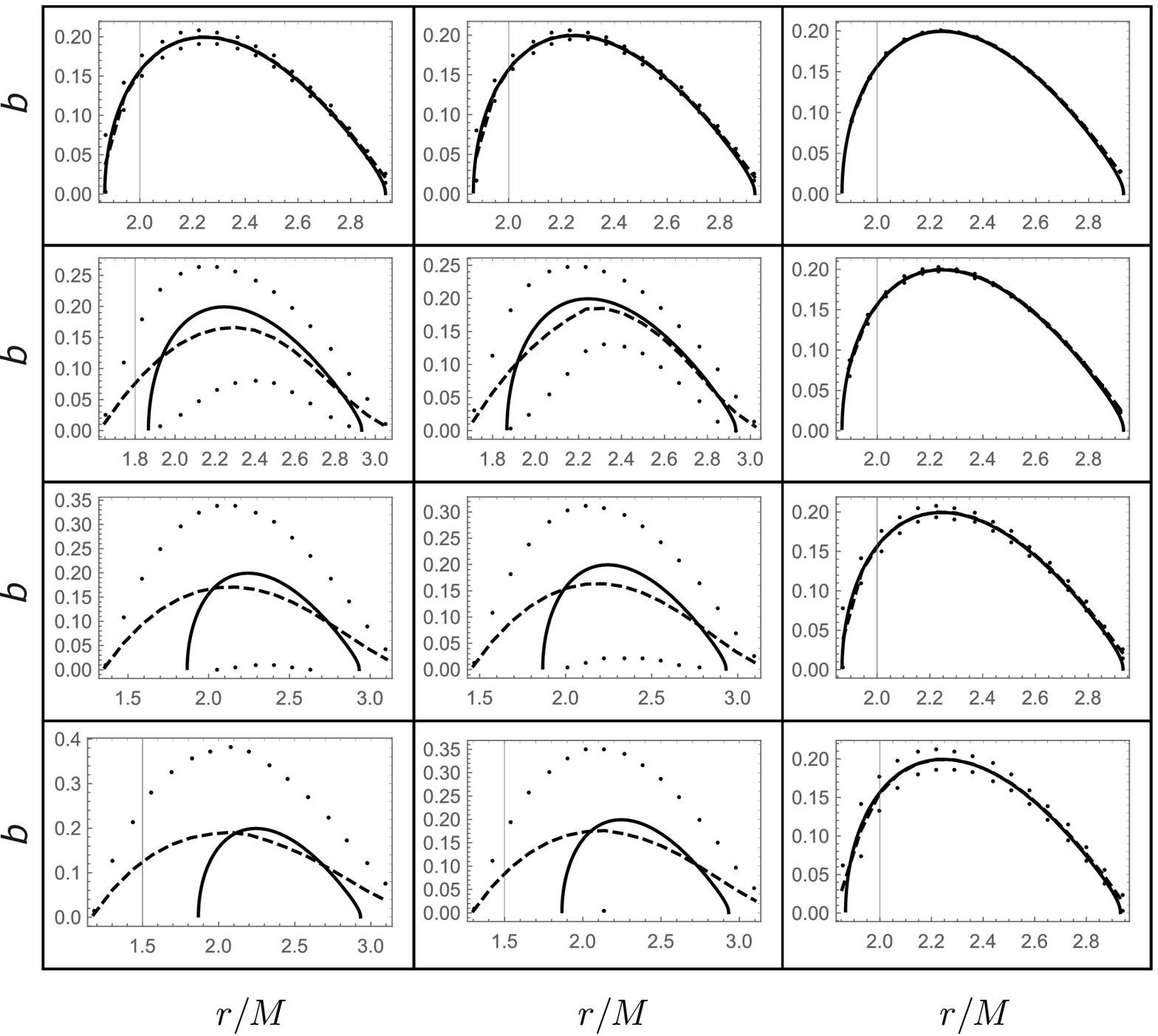}
	\caption{The radial profile of the local trapping efficiency coefficient $b$ is given for for the first-order Hartle--Thorne--Tolman VII object with 
		$R/M=3.1$ is plotted for several values of $j$ and $\theta$. 
		The curves and a choice of $j$ and $\theta$ are the same as in Fig.~\ref{loct27}.}
	\label{loct31}
\end{figure}

\begin{figure}[ht]
	\centering
	\includegraphics[width=0.90\hsize]{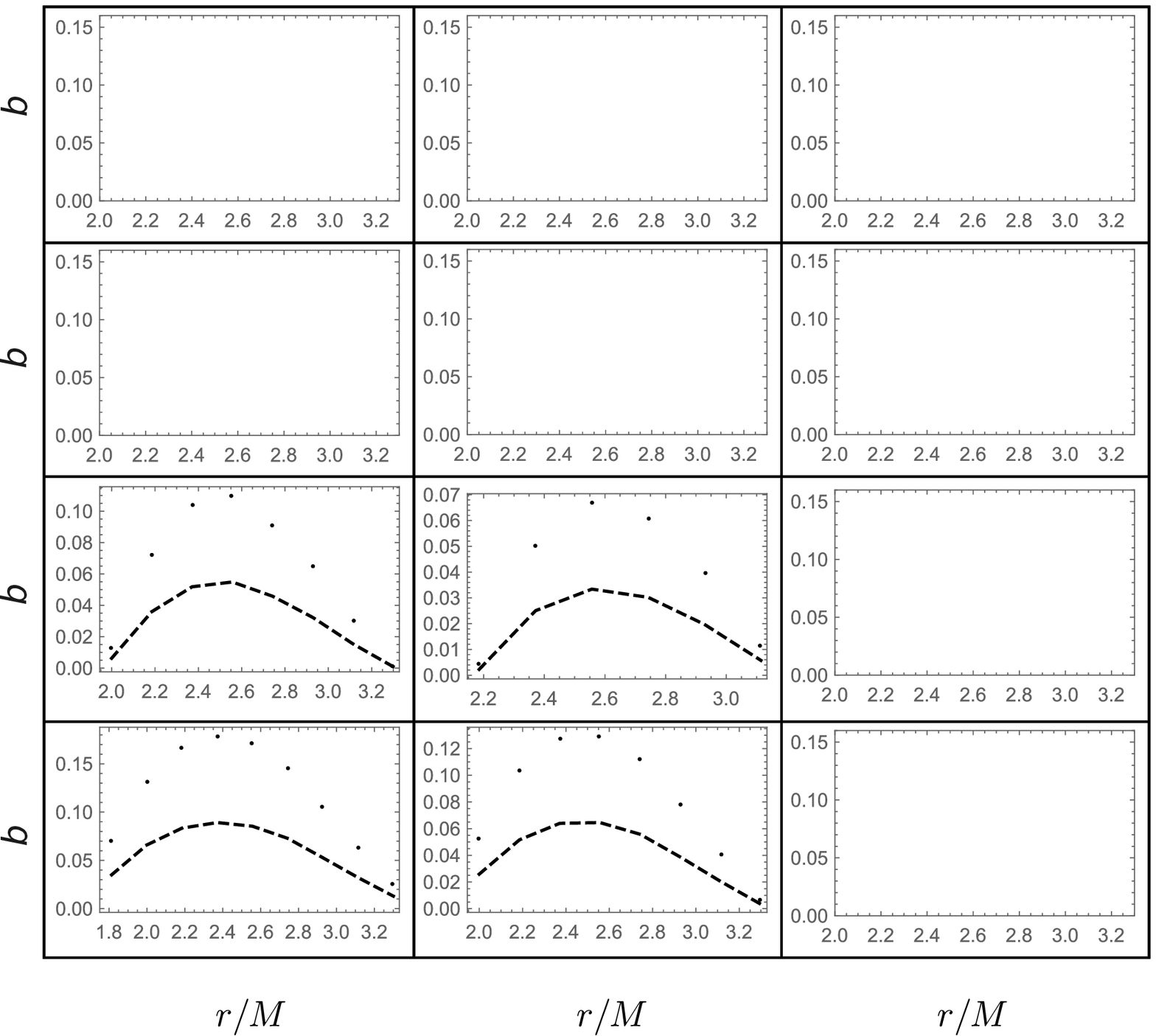}
	\caption{The radial profile of the local trapping efficiency coefficient $b$ is given for for the first-order Hartle--Thorne--Tolman VII object with 
		$R/M=3.3$ is plotted for several values of $j$ and $\theta$. 
		The curves and a choice of $j$ and $\theta$ are the same as in Fig.~\ref{loct27}.}
	\label{loct33}
\end{figure}

At a given point of the compact object, determined by coordinates $r,\theta$, we define the local trapping efficiency coefficient $b$ as the ratio of the number of the neutrinos emitted from this point and trapped by the object, $N_b$, to the number of neutrinos totally produced at this point $N_p$ (for details see \cite{Stu-Tor-Hle:2009:CQG}). Due to the assumed isotropy of the radiation emitted by the local source, the local trapping coefficient is given by the ratio of the surface of the trapping cone $S_\mathrm{tr}$ to the total area $S_{tot}=4\pi$, i.e., purely by the geometry of the spacetime. Therefore,    
\begin{equation}
b(r,\theta;J,R)\equiv\frac{\mathrm{N_b(r)}}{\mathrm{N_p(r)}} = \frac{S_\mathrm{tr}}{4\pi}. 
\end{equation}
We use this procedure for all radii $r$ relevant for the trapping, i.e., $r \geq r_\mathrm{b(u)}$, see Fig.~\ref{f:vef}, where $r_\mathrm{b(u)}(\theta,R/M,j)$ is the radius where the trapping begins. The radial profiles of the local trapping efficiency coefficient are presented in figures \ref{loct27} - \ref{loct33}. For comparison we illustrate also the analytical solution obtained for the spherically symmetric Tolman VII spacetime in \cite{Stu-Vrb-Hla:2021:sub1} (solid line). To clearly demonstrate influence of the object rotation, we give separately the trapping coefficient for the co-rotating and counter-rotating neutrinos, splitting thus the coefficient $b$ into two complementary parts denoted as $b_+$ and $b_-$. Therefore, $b_+$ denotes the local trapping efficiency coefficient for co-rotating null geodesics represented by the right part of the trapping cones (containing $\beta=\pi/2$), while $b_-$ denotes the coefficient for counter-rotating null geodesics corresponding to the left part of the trapping cones (containing angle $\beta=3\pi/2$). We thus define 
\begin{equation}
             b_{+}(r,\theta;j,R) = \frac{S_\mathrm{tr+}}{2\pi} ,  b_{-}(r,\theta;j,R) = \frac{S_\mathrm{tr-}}{2\pi} . 
\end{equation}
In our figures, $b_-$ tends to be located above the analytical function representing the case of the spherically symmetric spacetime (if it exists) and $b_+$ tends to be located under this function.


We can see that generally the local trapping in the first-order Hartle--Thorne--Tolman VII spacetimes is lower in comparison with those of the spherically symmetric Tolman VII spacetimes, with exception of the deepest regions of the trapping that usually reach smaller radii in the rotating spacetimes (especially for the counter-rotating null geodesics). The extension of the trapping region in the rotating spacetimes, in comparison with the related spherical internal Tolman VII spacetimes, increases with increasing spacetime parameter $R/M$. As expected, in the first-order Hartle--Thorne--Tolman VII spacetimes with $R/M > 3.202$, the trapping effect can be relevant for the counter-rotating null geodesics only, and we observe the trapping only for sufficiently high rotation parameters, $j > 0.1$.

\subsection{Neutrino production}

To treat the efficiency of the neutrino trapping in the first-order Hartle--Thorne--Tolman VII spacetimes in the global sense reflecting both the cooling process and the total neutrino luminosity of the object, we have to reasonably estimate the production rate of neutrinos in the interior of the Tolman VII spacetimes. 

The neutrino production is fully governed by detailed structure of the compact object (e.g., a neutron star) and reflects all the physical complexities of the object interior. As the object is still considered to be spherically symmetric, given by the Tolman VII internal geometry, it is enough to consider only the dependence on the radius, not the latitude. The local neutrino production rate $I(r)$ is determined by the relation 
\begin{equation}
                 \pazocal{I}(r\{A\}) = \frac{dN(r\{A\})}{d\tau} , 
\end{equation}
where $dN$ is number of neutrino producing interactions at radius $r$ in element of proper time $d\tau$ of the static observer located at the radius; {A} denotes the full set of quantities determining the neutrino production rate considered at the given radius. The interactions number reads 
\begin{equation}
                 \mathrm{d}N(r\{A\}) = dn(r) \Gamma(r) \mathrm{d}V(r) , 
\end{equation}
where $\mathrm{d}n$ denotes the number density of particles determining the neutrino production, $\Gamma$ denotes the neutrino production rate (governed by the temperature at given radius) and $\mathrm{d}V$ denotes the proper volume element at given radius. The quantities $\mathrm{d}n$ and $\Gamma$ are thus determined by physical conditions in the matter of the compact object, while $\mathrm{d}V$ is determined by the geometry. The production rate of neutrinos is a very complex function of radius, governed by all physical complexities of the compact object. However, it is not necessary to consider all the physical details of the compact object and related consequences on the neutrino production rate in the case of the Tolman VII solution, as it represents only a rough approximation given by the fixed energy density radial profile. The influence of the modification of the energy density profile in the Tolman VII solution can be sufficiently represented by assumption that the neutrino production rate is determined by the energy density profile, as in the previous studies of the trapping effect in the internal Schwarzschild spacetimes \cite{Stu-Kot:2009:GRG,Stu-Hla-Urb:2012:GRaG}. Naturally, the energy density radial profile includes in an implicit way the effect of temperature of the matter \cite{Web:1999:BOOK}. We thus consider the neutrino production rate in the form 
\begin{equation}
                 \pazocal{I}(r) = \frac{\mathrm{d}\pazocal{N}(r)}{d\tau} \sim \rho(r) .  
\end{equation}
In accord with \cite{Stu-Tor-Hle:2009:CQG}, we also apply the assumption of locally isotropic neutrino radiation implying that the efficiency of the trapping effect is given by the spacetime geometry only, namely by the local trapping efficiency coefficient $b$. 

Due to the time-delay factor the neutrino production rate related to the distant static observers can be expressed as  
\begin{equation}
                 I(r) = \frac{\mathrm{d}\pazocal{N}(r)}{\mathrm{d} t} = \pazocal{I}e^{\Phi(r)/2} ,   
\end{equation}
and the number of neutrinos generated at the proper volume element $\mathrm{d}V$ at given radius and in unit of time of distant static observers, i.e., the local neutrino production rate, can be expressed in the form 
\begin{eqnarray}
	\mathrm{d}N_{p}(r) &=& I(r)\mathrm{d}V(r) = 4\pi \pazocal{I}(r) e^{\Phi(r)/2} e^{\Psi(r)/2} r^2 \mathrm{d}r\nonumber \\
	&\sim& 4\pi \rho(r) e^{\Phi(r)/2} e^{\Psi(r)/2} r^2 \mathrm{d}r \nonumber \\
	&=& 4\pi \rho_{c}(1 - \frac{r^2}{R^2}) e^{\Phi(r)/2} e^{\Psi(r)/2} r^2 \mathrm{d}r.
\end{eqnarray}
The global neutrino production rate is determined by 
\begin{eqnarray}
	N_{p} &=& 4\pi \int_{0}^{R} \pazocal{I}(r) e^{\Phi(r)/2} e^{\Psi(r)/2} r^2 \, \mathrm{d}r \nonumber \\ 
	&=& 4\pi \int_{0}^{R}
	\rho_{c}(1 - \frac{r^2}{R^2}) e^{\Phi(r)} e^{\Psi(r)} r^2 \, \mathrm{d}r , 
\end{eqnarray}
while the global rate of the neutrino trapping is under the assumption of the isotropy of the emitted neutrino flow given by 
\begin{eqnarray}
	N_{p} &=& 4\pi \int_{0}^{\pi/2} \int_{r_\mathrm{b(u)}}^{\min\{R,r_\mathrm{c(u)}\}} b(r,\theta) \pazocal{I}(r) e^{\Phi(r)/2} e^{\Psi(r)/2} r^2\,  
	\mathrm{d}r\, \mathrm{d}\theta \nonumber \\
	&=&4\pi \int_{0}^{\pi/2} \int_{r_\mathrm{b(u)}}^{\min\{R,r_\mathrm{c(u)}\}} b(r,\theta) \rho_{c}\left(1 - \frac{r^2}{R^2}\right) e^{\Phi(r)/2}\, e^{\Psi(r)/2} r^2\, \mathrm{d}r\, \mathrm{d}\theta.
\end{eqnarray}

\subsection{Global trapping}

The coefficient of the global trapping reflects the trapping phenomenon integrated across the whole trapping region, related to the whole radiating object. We thus consider the amount of neutrinos radiated along null geodesics by the whole object in the unit time of distant static observers, and determine the part of these radiated neutrinos that remains trapped by the radiating object. Details of the derivation of the global trapping coefficient are presented in \cite{Stu-Tor-Hle:2009:CQG,Vrb-Urb-Stu:2020:EPJC}, and we apply them in our paper using again the basic assumption that the locally defined radiation intensity is proportional to the energy density of the object, being thus distributed quadratically across whole the Tolman VII object. 

The global trapping effects are thus reflected by the global trapping efficiency coefficient $\pazocal{B}$ defined by the relation \cite{Stu-Tor-Hle:2009:CQG}
\begin{equation}
\pazocal{B}\equiv\frac{N_b}{N_p}=\frac{\int_{0}^{R}\int^{\pi}_{0}\int^{2\pi}_{0}f^{-1}(r)b(r,\theta) \, r^2 \, \mathrm{d}\varphi \,\mathrm{d}\theta \, \mathrm{d}r}{\int_{0}^{R}\int_{0}^{\pi}\int_{0}^{2\pi}f^{-1}(r) \, r^2 \, \mathrm{d}\varphi \,\mathrm{d}\theta \, \mathrm{d}r},
\end{equation}
where we use the radial metric function $f(r)$ defined in Eq.~(\ref{e:f}).

But we can simplify the integration process due to the symmetries of the first-order Hartle--Thorne--Tolman VII spacetime. The metric coefficients are independent of $\varphi$, and the results of integration in the top and lower hemisphere are the same. Also, limits in the radial direction can be shrunk by using knowledge of the position of $ r_\mathrm{b(u)}(r,\theta;R,J,\lambda)$ determining the limits of integration of the trapping effect. So the global trapping efficiency coefficient is presented in the form  
\begin{equation}
\pazocal{B}=\frac{\int_{r_\mathrm{b(u)}}^{R}\int^{\pi/2}_{0}f^{-1}(r)b(r,\theta) \, r^2 \, \mathrm{d}\theta \, \mathrm{d}r}{\int_{0}^{R}\int_{0}^{\pi/2}f^{-1}(r) \, r^2 \, \mathrm{d}\theta \, \mathrm{d}r},
\end{equation}
for the unstable null geodesic located outside the object ($R/M<3$) or
\begin{equation}
	\pazocal{B}=\frac{\int_{r_\mathrm{b(u)}}^{r_\mathrm{c(u)}}\int^{\pi/2}_{0}f^{-1}(r)b(r,\theta) \, r^2 \, \mathrm{d}\theta \, \mathrm{d}r}{\int_{0}^{R}\int_{0}^{\pi/2}f^{-1}(r) \, r^2 \, \mathrm{d}\theta \, \mathrm{d}r},
\end{equation}
for the unstable null geodesic inside the object. 

\begin{figure}[ht]
\centering
\includegraphics[width=0.7\hsize]{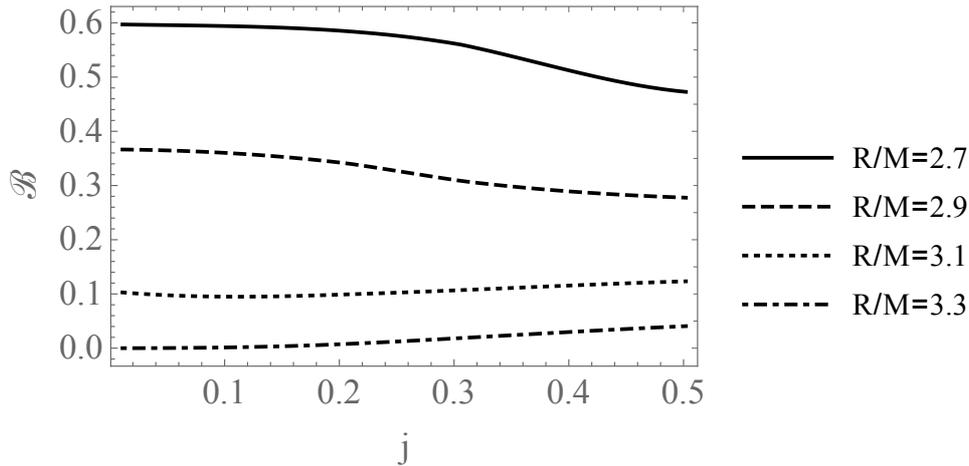}
\caption{The comparison of the global trapping efficiency coefficient $\pazocal{B}$ for various parameter $R/M$.}
\label{f:glob}
\end{figure}

Figure \ref{f:glob} shows how the global trapping efficiency coefficient $\pazocal{B}$ depends on the rotation parameter $j$ for fixed characteristic values of the inverse compactness $R/M$ of the object. \footnote{In order to indicate possible behaviour of the trapping effects for the standard internal Hartle--Thorne--Tolman VII spacetimes, we consider here also relatively large values of the rotation parameter $j$ rising up to $j=0.5$. Note that $j=0.7$ is considered to be the limiting value acceptable for the Hartle--Thorne spacetimes taken to the quadratic approximation in the angular velocity $\Omega$ \cite{Urb-Mil-Stu:2013:MNRAS}.} Notice that for the low value of the inverse compactness parameter, namely $R/M=2.7$ and  $R/M=2.9$, the global efficiency parameter $\pazocal{B}(j)$ decreases monotonically with increasing $j$. On the other hand, we observe decrease of $\pazocal{B}$ up to $j \sim 0.2$ and increase at $j>0.2$ for the large value of $R/M=3.1$, when the trapping is impossible in the spherical internal Schwarzschild spacetimes. For the first-order Hartle--Thorne--Tolman VII object with $R/M=3.3$, the trapping occurs above $j=0.1$ and increases with increasing rotation parameter $j$, but exclusively for the counter-rotating null geodesics. Nevertheless, we have to say that detailed calculations in the full Hartle--Thorne geometry are necessary to obtain precise description of the trapping phenomenon for rotation parameters larger than $j = 0.1$ when the linearity assumption of the Hartle--Thorne metric in $\Omega$ could be violated. 

\FloatBarrier

\section{Conclusions}

In our introductory study on the role of the rotation on the phenomenon of trapping of null geodesics that could be relevant for motion of neutrinos in the interior of neutron stars, we have used strongest simplification of the first-order (linearized) Hartle--Thorne spacetime with uniformly distributed energy density of matter \cite{Vrb-Urb-Stu:2020:EPJC}, in order to obtain simple and easily tractable results. 

Here we have discussed the role of the first-order rotational phenomena of the trapping in the Hartle--Thorne--Tolman VII spacetimes where the energy density demonstrates quadratic radial profile, making such simple exact solutions of Einstein equations close to realistic models of neutron stars \cite{Jia-Yag:2019:PRD,Pos-Hla-Stu:2021:sub}. We have found that the rotation of the Hartle--Thorne--Tolman VII objects enhances the trapping effects even more efficiently than in the case of internal Scwarzshchild spacetimes, enabling them even at objects with $R/M \sim 3.3$ which are close to those in observed neutron stars. 

We believe that even such a simplification enables to find the basic characteristics of the influence of the rotation of radiating compact objects on the effect of trapping in their interior. For these purposes we are considering also values of the dimensionless rotation parameter $j$ overcoming the values of safe validity of the linear, first-order approximation that cannot be higher than $j \sim 0.2$. Nevertheless, we expect that the results obtained for values of $j > 0.2$ could indicate relevant signatures of realistic effects even in the linear approximation. 

Our results related to the local effects indicate much stronger trapping coefficient of the counter-rotating null geodesics in comparison with the corotating ones and even with the trapping coefficient in the internal Tolman VII spacetimes with the same parameter $R/M$. In the first-order Hartle--Thorne--Tolman VII spacetimes the region of trapping is always larger than in the related spherically symmetric internal Tolman VII spacetimes having the same parameter $R/M$, this difference increases with increasing $R/M$. 

The trapping of the counter-rotating null geodesics has been demonstrated even for $R/M>3.202$, i.e., with values forbidden for the trapping spherical Tolman VII spacetimes. The trapping was relevant even for objects with $R/M=3.3$, but only for the rotation parameter starting at $j \sim 0.2$ when validity of the linear approximation of the Hartle--Thorne metric starts to be limited. Therefore, more detailed models based on the complete Hartle--Thorne geometry are necessary to confirm the possibility to overcome the limit for trapping effects at the radius $R=3.2M$, and to make precise mapping of the trapping effect for values of the rotation parameter $j>0.2$ when some relevance of the linear approximation is expected, up to the limiting value of $j=0.5$ when the full Hartle--Thorne metric  \cite{Urb-Mil-Stu:2013:MNRAS} is necessary that fully reflects the second-order effect of the angular velocity. 

We can conclude that our results indicate relevance of the neutrino trapping in realistic neutron stars that can be well approximated by the exact Tolman VII solution of the Einstein equations \cite{Jia-Yag:2019:PRD,Pos-Hla-Stu:2021:sub}. The neutrinos bounded in the trapping region will be eventually re-scattered and after such a process they could leave the trapping region or continue motion inside the region, in dependence on accidental condition of the scatter process. Of course, in the neutrino dominated cooling of the object the scattering of trapped neutrinos causes increase of the temperature inside the trapping region in comparison with the internal regions from which the neutrinos escape directly. The trapping region can finally have higher temperature than the interior, and some other agents than neutrinos could cause an inflow of heat from the overheated external trapping region to the interior influencing the structure of the extremely compact object. In such a special process of the cooling of the extremely compact objects we could even expect formation of some self-organized structure in vicinity of the boundary of the trapping region.


\section*{acknowledgments}
J.V. and Z.S. acknowledge the institutional support of the Institute of Physics, Silesian University in Opava. J.V. were supported by the Czech Grant No. LTC18058.

\bibliographystyle{abbrvnat}
\bibliography{references}

\providecommand{\noopsort}[1]{}\providecommand{\singleletter}[1]{#1}%
\begin{thebibliography}{48}
\providecommand{\natexlab}[1]{#1}
\providecommand{\url}[1]{\texttt{#1}}
\expandafter\ifx\csname urlstyle\endcsname\relax
  \providecommand{\doi}[1]{doi: #1}\else
  \providecommand{\doi}{doi: \begingroup \urlstyle{rm}\Url}\fi

\bibitem[{Abramowicz} et~al.(1993){Abramowicz}, {Miller}, and
  {Stuchl{\'\i}k}]{Abr-Mil-Stu:1993:PRD}
M.~A. {Abramowicz}, J.~C. {Miller}, and Z.~{Stuchl{\'\i}k}.
\newblock {Concept of radius of gyration in general relativity}.
\newblock \emph{Phys. Rev. D}, 47\penalty0 (4):\penalty0 1440--1447, Feb. 1993.
\newblock \doi{10.1103/PhysRevD.47.1440}.

\bibitem[{Abramowicz} et~al.(1997){Abramowicz}, {Andersson}, {Bruni}, {Ghosh},
  and {Sonego}]{Abr-And-Bru:1997:CQG}
M.~A. {Abramowicz}, N.~{Andersson}, M.~{Bruni}, P.~{Ghosh}, and S.~{Sonego}.
\newblock {LETTER TO THE EDITOR: Gravitational waves from ultracompact stars:
  the optical geometry view of trapped modes}.
\newblock \emph{Classical and Quantum Gravity}, 14\penalty0 (12):\penalty0
  L189--L194, Dec. 1997.
\newblock \doi{10.1088/0264-9381/14/12/002}.

\bibitem[{Abramowicz} et~al.(2003){Abramowicz}, {Almergren}, {Kluzniak}, and
  {Thampan}]{Abr-Alm-Klu:2003:arx}
M.~A. {Abramowicz}, G.~J.~E. {Almergren}, W.~{Kluzniak}, and A.~V. {Thampan}.
\newblock {The Hartle-Thorne circular geodesics}.
\newblock \emph{arXiv e-prints}, art. gr-qc/0312070, Dec. 2003.

\bibitem[{Barack} and {et al.}(2019)]{Bar-etal:2019:CQG}
L.~{Barack} and {et al.}
\newblock {Black holes, gravitational waves and fundamental physics: a
  roadmap}.
\newblock \emph{Classical and Quantum Gravity}, 36\penalty0 (14):\penalty0
  143001, July 2019.
\newblock \doi{10.1088/1361-6382/ab0587}.

\bibitem[{Bardeen} et~al.(1972){Bardeen}, {Press}, and
  {Teukolsky}]{Bar-Pre-Teu:1972:ApJ}
J.~M. {Bardeen}, W.~H. {Press}, and S.~A. {Teukolsky}.
\newblock {Rotating Black Holes: Locally Nonrotating Frames, Energy Extraction,
  and Scalar Synchrotron Radiation}.
\newblock \emph{Astrophys. J.}, 178:\penalty0 347--370, Dec. 1972.
\newblock \doi{10.1086/151796}.

\bibitem[{B{\"o}hmer}(2004)]{Boh:2004:GRG}
C.~G. {B{\"o}hmer}.
\newblock {Eleven Spherically Symmetric Constant Density Solutions with
  Cosmological Constant}.
\newblock \emph{General Relativity and Gravitation}, 36\penalty0 (5):\penalty0
  1039--1054, May 2004.
\newblock \doi{10.1023/B:GERG.0000018088.69051.3b}.

\bibitem[{Cardoso} et~al.(2009){Cardoso}, {Miranda}, {Berti}, {Witek}, and
  {Zanchin}]{Car-Mir-Ber:2009:PRD}
V.~{Cardoso}, A.~S. {Miranda}, E.~{Berti}, H.~{Witek}, and V.~T. {Zanchin}.
\newblock {Geodesic stability, Lyapunov exponents, and quasinormal modes}.
\newblock \emph{Phys. Rev. D}, 79\penalty0 (6):\penalty0 064016, Mar. 2009.
\newblock \doi{10.1103/PhysRevD.79.064016}.

\bibitem[{Chandrasekhar}(1983)]{cha:1983:BOOK}
S.~{Chandrasekhar}.
\newblock \emph{{The mathematical theory of black holes}}.
\newblock Oxford/New York, Clarendon Press/Oxford University Press, 1983.

\bibitem[{Felice}(1968)]{Fel:1968:NCBS}
F.~{Felice}.
\newblock {Equatorial geodesic motion in the gravitational field of a rotating
  source}.
\newblock \emph{Nuovo Cimento B Serie}, 57\penalty0 (2):\penalty0 351--388,
  Oct. 1968.
\newblock \doi{10.1007/BF02710207}.

\bibitem[{Glendenning}(2000)]{Gle:2000:BOOK}
N.~K. {Glendenning}, editor.
\newblock \emph{{Compact stars : nuclear physics, particle physics, and general
  relativity}}, 2000.

\bibitem[{Hartle} and {Thorne}(1968)]{Har-Tho:1968:ApJ}
J.~B. {Hartle} and K.~S. {Thorne}.
\newblock {Slowly Rotating Relativistic Stars. II. Models for Neutron Stars and
  Supermassive Stars}.
\newblock \emph{Astrophys. J.}, 153:\penalty0 807--+, Sept. 1968.
\newblock \doi{10.1086/149707}.

\bibitem[{Hensh} and {Stuchl{\'\i}k}(2019)]{Hen-Stu:2019:GRQC}
S.~{Hensh} and Z.~{Stuchl{\'\i}k}.
\newblock {Anisotropic Tolman VII solution by gravitational decoupling}.
\newblock \emph{European Physical Journal C}, 79\penalty0 (10):\penalty0 834,
  Oct. 2019.
\newblock \doi{10.1140/epjc/s10052-019-7360-9}.

\bibitem[{Hod}(2018)]{Hod:2018:PRD}
S.~{Hod}.
\newblock {Lower bound on the compactness of isotropic ultracompact objects}.
\newblock \emph{\prd}, 97\penalty0 (8):\penalty0 084018, Apr. 2018.
\newblock \doi{10.1103/PhysRevD.97.084018}.

\bibitem[{Jiang} and {Yagi}(2019)]{Jia-Yag:2019:PRD}
N.~{Jiang} and K.~{Yagi}.
\newblock {Improved analytic modeling of neutron star interiors}.
\newblock \emph{\prd}, 99\penalty0 (12):\penalty0 124029, June 2019.
\newblock \doi{10.1103/PhysRevD.99.124029}.

\bibitem[{Jiang} and {Yagi}(2020)]{Jia-Yag:2020:PRD}
N.~{Jiang} and K.~{Yagi}.
\newblock {Analytic I-Love-C relations for realistic neutron stars}.
\newblock \emph{\prd}, 101\penalty0 (12):\penalty0 124006, June 2020.
\newblock \doi{10.1103/PhysRevD.101.124006}.

\bibitem[{Konoplya} and {Stuchl{\'\i}k}(2017)]{Kon-Stu:2017:PLB}
R.~A. {Konoplya} and Z.~{Stuchl{\'\i}k}.
\newblock {Are eikonal quasinormal modes linked to the unstable circular null
  geodesics?}
\newblock \emph{Physics Letters B}, 771:\penalty0 597--602, Aug. 2017.
\newblock \doi{10.1016/j.physletb.2017.06.015}.

\bibitem[{Konoplya} et~al.(2019){Konoplya}, {Posada}, {Stuchl{\'\i}k}, and
  {Zhidenko}]{Kon-Pos-Stu:2019:PRD}
R.~A. {Konoplya}, C.~{Posada}, Z.~{Stuchl{\'\i}k}, and A.~{Zhidenko}.
\newblock {Stable Schwarzschild stars as black-hole mimickers}.
\newblock \emph{\prd}, 100\penalty0 (4):\penalty0 044027, Aug. 2019.
\newblock \doi{10.1103/PhysRevD.100.044027}.

\bibitem[{Miller}(1977)]{Mil:1977:MNRAS}
J.~C. {Miller}.
\newblock {Quasi-stationary gravitational collapse of slowly rotating bodies in
  general relativity}.
\newblock \emph{Mon. Not. R. Astron. Soc.}, 179:\penalty0 483--498, May 1977.
\newblock \doi{10.1093/mnras/179.3.483}.

\bibitem[{Misner} et~al.(1973){Misner}, {Thorne}, and
  {Wheeler}]{Mis-Tho-Whe:1973:BOOK}
C.~W. {Misner}, K.~S. {Thorne}, and J.~A. {Wheeler}.
\newblock \emph{{Gravitation}}.
\newblock 1973.

\bibitem[{Neary} et~al.(2001){Neary}, {Ishak}, and
  {Lake}]{Nea-Ish-Lak:2001:PRD}
N.~{Neary}, M.~{Ishak}, and K.~{Lake}.
\newblock {Tolman type VII solution, trapped null orbits, and w-modes}.
\newblock \emph{\prd}, 64\penalty0 (8):\penalty0 084001, Oct. 2001.
\newblock \doi{10.1103/PhysRevD.64.084001}.

\bibitem[{Novotn{\'y}} et~al.(2017){Novotn{\'y}}, {Hlad{\'\i}k}, and
  {Stuchl{\'\i}k}]{Nov-Hla-Stu:2017:PRD}
J.~{Novotn{\'y}}, J.~{Hlad{\'\i}k}, and Z.~{Stuchl{\'\i}k}.
\newblock {Polytropic spheres containing regions of trapped null geodesics}.
\newblock \emph{Phys. Rev. D}, 95\penalty0 (4):\penalty0 043009, Feb. 2017.
\newblock \doi{10.1103/PhysRevD.95.043009}.

\bibitem[{Ovalle} et~al.(2019){Ovalle}, {Posada}, and
  {Stuchl{\'\i}k}]{Ova-Pos-Stu:2019:CQG}
J.~{Ovalle}, C.~{Posada}, and Z.~{Stuchl{\'\i}k}.
\newblock {Anisotropic ultracompact Schwarzschild star by gravitational
  decoupling}.
\newblock \emph{Classical and Quantum Gravity}, 36\penalty0 (20):\penalty0
  205010, Oct. 2019.
\newblock \doi{10.1088/1361-6382/ab4461}.

\bibitem[{Peng}(2020)]{Pen:2020:ARX}
Y.~{Peng}.
\newblock {Upper bounds on the compactness at the innermost light ring of
  anisotropic horizonless spheres}.
\newblock \emph{European Physical Journal C}, 80\penalty0 (8):\penalty0 755,
  Aug. 2020.
\newblock \doi{10.1140/epjc/s10052-020-8358-z}.

\bibitem[{Posada} and {Chirenti}(2019)]{Pos-Chi:2019:CQG}
C.~{Posada} and C.~{Chirenti}.
\newblock {On the radial stability of ultra-compact Schwarzschild stars beyond
  the Buchdahl limit}.
\newblock \emph{Classical and Quantum Gravity}, 36\penalty0 (6):\penalty0
  065004, Mar. 2019.
\newblock \doi{10.1088/1361-6382/ab0526}.

\bibitem[{Posada} et~al.(2021){Posada}, {Hlad{\'\i}k}, and
  {Stuchl{\'\i}k}]{Pos-Hla-Stu:2021:sub}
C.~{Posada}, J.~{Hlad{\'\i}k}, and Z.~{Stuchl{\'\i}k}.
\newblock {Dynamical stability of the modified Tolman VII solution}.
\newblock \emph{arXiv e-prints}, art. arXiv:2103.12867, Mar. 2021.

\bibitem[{Randall} and {Sundrum}(1999)]{Ran-Sun:1999:PRL}
L.~{Randall} and R.~{Sundrum}.
\newblock {An Alternative to Compactification}.
\newblock \emph{Physical Review Letters}, 83:\penalty0 4690--4693, Dec. 1999.
\newblock \doi{10.1103/PhysRevLett.83.4690}.

\bibitem[{Schee} and {Stuchl{\'{\i}}k}(2009)]{Sch-Stu:2009:IJMPD}
J.~{Schee} and Z.~{Stuchl{\'{\i}}k}.
\newblock {Optical Phenomena in the Field of Braneworld Kerr Black Holes}.
\newblock \emph{International Journal of Modern Physics D}, 18:\penalty0
  983--1024, 2009.
\newblock \doi{10.1142/S0218271809014881}.

\bibitem[{Schwarzschild}(1916)]{Sch:1916:AkadW}
K.~{Schwarzschild}.
\newblock {On the Gravitational Field of a Mass Point According to Einstein's
  Theory}.
\newblock \emph{Abh. Konigl. Preuss. Akad. Wissenschaften Jahre 1906,92,
  Berlin,1907}, 1916:\penalty0 189--196, Jan. 1916.

\bibitem[{Shapiro} and {Teukolsky}(1986)]{Sha-Stu-Teu:1986:book}
S.~L. {Shapiro} and S.~A. {Teukolsky}.
\newblock \emph{{Black Holes, White Dwarfs and Neutron Stars: The Physics of
  Compact Objects}}.
\newblock 1986.

\bibitem[{Stuchl{\'\i}k}(2000)]{Stu:2000:APS}
Z.~{Stuchl{\'\i}k}.
\newblock {Spherically Symmetric Static Configurations of Uniform Density in
  Spacetimes with a Non-Zero Cosmological Constant}.
\newblock \emph{Acta Physica Slovaca}, 50\penalty0 (2):\penalty0 219--228, Apr.
  2000.

\bibitem[{Stuchl{\'{\i}}k}(2005)]{Stu:2005:MPLA}
Z.~{Stuchl{\'{\i}}k}.
\newblock {Influence of the RELICT Cosmological Constant on Accretion Discs}.
\newblock \emph{Modern Physics Letters A}, 20:\penalty0 561--575, 2005.
\newblock \doi{10.1142/S0217732305016865}.

\bibitem[{Stuchl{\'{\i}}k} and {Kotrlov{\'a}}(2009)]{Stu-Kot:2009:GRG}
Z.~{Stuchl{\'{\i}}k} and A.~{Kotrlov{\'a}}.
\newblock {Orbital resonances in discs around braneworld Kerr black holes}.
\newblock \emph{General Relativity and Gravitation}, 41:\penalty0 1305--1343,
  June 2009.
\newblock \doi{10.1007/s10714-008-0709-2}.

\bibitem[{Stuchl{\'{\i}}k} and {Schee}(2010)]{Stu-Sch:2010:CQG}
Z.~{Stuchl{\'{\i}}k} and J.~{Schee}.
\newblock {Appearance of Keplerian discs orbiting Kerr superspinars}.
\newblock \emph{Classical and Quantum Gravity}, 27\penalty0 (21):\penalty0
  215017, Nov. 2010.
\newblock \doi{10.1088/0264-9381/27/21/215017}.

\bibitem[{Stuchl{\'\i}k} and {Schee}(2019)]{Stu-Sch:2019:EPJC}
Z.~{Stuchl{\'\i}k} and J.~{Schee}.
\newblock {Shadow of the regular Bardeen black holes and comparison of the
  motion of photons and neutrinos}.
\newblock \emph{European Physical Journal C}, 79\penalty0 (1):\penalty0 44,
  Jan. 2019.
\newblock \doi{10.1140/epjc/s10052-019-6543-8}.

\bibitem[{Stuchl{\'\i}k} et~al.(2009){Stuchl{\'\i}k}, {T{\"o}r{\"o}k},
  {Hled{\'\i}k}, and {Urbanec}]{Stu-Tor-Hle:2009:CQG}
Z.~{Stuchl{\'\i}k}, G.~{T{\"o}r{\"o}k}, S.~{Hled{\'\i}k}, and M.~{Urbanec}.
\newblock {Neutrino trapping in extremely compact objects: I. Efficiency of
  trapping in the internal Schwarzschild spacetimes}.
\newblock \emph{Classical and Quantum Gravity}, 26\penalty0 (3):\penalty0
  035003, Feb. 2009.
\newblock \doi{10.1088/0264-9381/26/3/035003}.

\bibitem[{Stuchl{\'{\i}}k} et~al.(2011){Stuchl{\'{\i}}k}, {Hlad{\'{\i}}k}, and
  {Urbanec}]{Stu-Hla-Urb:2011:GRG}
Z.~{Stuchl{\'{\i}}k}, J.~{Hlad{\'{\i}}k}, and M.~{Urbanec}.
\newblock {Neutrino trapping in braneworld extremely compact stars}.
\newblock \emph{General Relativity and Gravitation}, 43:\penalty0 3163--3190,
  Nov. 2011.
\newblock \doi{10.1007/s10714-011-1229-z}.

\bibitem[{Stuchl{\'\i}k} et~al.(2012){Stuchl{\'\i}k}, {Hlad{\'\i}k}, {Urbanec},
  and {T{\"o}r{\"o}k}]{Stu-Hla-Urb:2012:GRaG}
Z.~{Stuchl{\'\i}k}, J.~{Hlad{\'\i}k}, M.~{Urbanec}, and G.~{T{\"o}r{\"o}k}.
\newblock {Neutrino trapping in extremely compact objects described by the
  internal Schwarzschild-(anti-)de Sitter spacetimes}.
\newblock \emph{General Relativity and Gravitation}, 44\penalty0 (6):\penalty0
  1393--1417, June 2012.
\newblock \doi{10.1007/s10714-012-1346-3}.

\bibitem[{Stuchl{\'{\i}}k} et~al.(2016){Stuchl{\'{\i}}k}, {Hled{\'{\i}}k}, and
  {Novotn{\'y}}]{Stu-Hle-Nov:2016:PRD}
Z.~{Stuchl{\'{\i}}k}, S.~{Hled{\'{\i}}k}, and J.~{Novotn{\'y}}.
\newblock {General relativistic polytropes with a repulsive cosmological
  constant}.
\newblock \emph{Phys. Rev. D}, 94\penalty0 (10):\penalty0 103513, Nov. 2016.
\newblock \doi{10.1103/PhysRevD.94.103513}.

\bibitem[{Stuchl{\'{\i}}k} et~al.(2017){Stuchl{\'{\i}}k}, {Schee}, {Toshmatov},
  {Hlad{\'{\i}}k}, and {Novotn{\'y}}]{Stu-Sch-Tos:2017:JCAP}
Z.~{Stuchl{\'{\i}}k}, J.~{Schee}, B.~{Toshmatov}, J.~{Hlad{\'{\i}}k}, and
  J.~{Novotn{\'y}}.
\newblock {Gravitational instability of polytropic spheres containing region of
  trapped null geodesics: a possible explanation of central supermassive black
  holes in galactic halos}.
\newblock \emph{JCAP}, 6:\penalty0 056, June 2017.
\newblock \doi{10.1088/1475-7516/2017/06/056}.

\bibitem[{Stuchl{\'\i}k} et~al.(2018){Stuchl{\'\i}k}, {Charbul{\'a}k}, and
  {Schee}]{Stu-Cha-Sch:2018:EPJC}
Z.~{Stuchl{\'\i}k}, D.~{Charbul{\'a}k}, and J.~{Schee}.
\newblock {Light escape cones in local reference frames of Kerr-de Sitter black
  hole spacetimes and related black hole shadows}.
\newblock \emph{European Physical Journal C}, 78\penalty0 (3):\penalty0 180,
  Mar. 2018.
\newblock \doi{10.1140/epjc/s10052-018-5578-6}.

\bibitem[{Stuchl{\'\i}k} et~al.(2020){Stuchl{\'\i}k}, {Kolo{\v{s}}},
  {Kov{\'a}{\v{r}}}, {Slan{\'y}}, and {Tursunov}]{Stu-Kol-Kov:2020:UNI}
Z.~{Stuchl{\'\i}k}, M.~{Kolo{\v{s}}}, J.~{Kov{\'a}{\v{r}}}, P.~{Slan{\'y}}, and
  A.~{Tursunov}.
\newblock {Influence of Cosmic Repulsion and Magnetic Fields on Accretion Disks
  Rotating around Kerr Black Holes}.
\newblock \emph{Universe}, 6\penalty0 (2):\penalty0 26, Jan. 2020.
\newblock \doi{10.3390/universe6020026}.

\bibitem[{Stuchl{\'\i}k} et~al.(2021){Stuchl{\'\i}k}, {Vrba}, {Hlad{\'\i}k},
  and {Posada}]{Stu-Vrb-Hla:2021:sub1}
Z.~{Stuchl{\'\i}k}, J.~{Vrba}, J.~{Hlad{\'\i}k}, and C.~{Posada}.
\newblock {Neutrino trapping in extremely compact Tolman VII spacetimes}.
\newblock \emph{European Physical Journal C}, submitted, Mar. 2021.

\bibitem[Tolman(1939)]{Tol:1939:PR}
R.~C. Tolman.
\newblock Static solutions of {E}instein's field equations for spheres of
  fluids.
\newblock \emph{Phys. Rev.}, 55:\penalty0 364, 1939.

\bibitem[{Toshmatov} et~al.(2018){Toshmatov}, {Stuchl{\'{\i}}k}, {Schee}, and
  {Ahmedov}]{Tos-Stu-Sch:2018:PRD}
B.~{Toshmatov}, Z.~{Stuchl{\'{\i}}k}, J.~{Schee}, and B.~{Ahmedov}.
\newblock {Electromagnetic perturbations of black holes in general relativity
  coupled to nonlinear electrodynamics}.
\newblock \emph{Phys. Rev. D}, 97\penalty0 (8):\penalty0 084058, Apr. 2018.
\newblock \doi{10.1103/PhysRevD.97.084058}.

\bibitem[{Toshmatov} et~al.(2019){Toshmatov}, {Stuchl{\'\i}k}, {Ahmedov}, and
  {Malafarina}]{Tos-Stu-Ahm:2019:GRQC}
B.~{Toshmatov}, Z.~{Stuchl{\'\i}k}, B.~{Ahmedov}, and D.~{Malafarina}.
\newblock {Relaxations of perturbations of spacetimes in general relativity
  coupled to nonlinear electrodynamics}.
\newblock \emph{\prd}, 99\penalty0 (6):\penalty0 064043, Mar. 2019.
\newblock \doi{10.1103/PhysRevD.99.064043}.

\bibitem[{Urbanec} et~al.(2013){Urbanec}, {Miller}, and
  {Stuchl{\'\i}k}]{Urb-Mil-Stu:2013:MNRAS}
M.~{Urbanec}, J.~C. {Miller}, and Z.~{Stuchl{\'\i}k}.
\newblock {Quadrupole moments of rotating neutron stars and strange stars}.
\newblock \emph{Monthly Notices of the RAS}, 433\penalty0 (3):\penalty0
  1903--1909, Aug. 2013.
\newblock \doi{10.1093/mnras/stt858}.

\bibitem[{Vrba} et~al.(2020){Vrba}, {Urbanec}, {Stuchl{\'\i}k}, and
  {Miller}]{Vrb-Urb-Stu:2020:EPJC}
J.~{Vrba}, M.~{Urbanec}, Z.~{Stuchl{\'\i}k}, and J.~C. {Miller}.
\newblock {Trapping of null geodesics in slowly rotating spacetimes}.
\newblock \emph{European Physical Journal C}, 80\penalty0 (11):\penalty0 1065,
  Nov. 2020.
\newblock \doi{10.1140/epjc/s10052-020-08642-z}.

\bibitem[Weber(1999)]{Web:1999:BOOK}
F.~Weber.
\newblock \emph{{Pulsars as astrophysical laboratories for nuclear and particle
  physics}}.
\newblock Jan. 1999.

\end{thebibliography}

\end{document}